\documentclass[review]{elsarticle}
\usepackage{geometry}
\newcommand*{\ARXIV}{}

\usepackage{comment}

\newcommand{\CFilesBib}{Common.Files.Bib}

\usepackage{amsmath,amssymb,amscd,latexsym,dsfont,wasysym,bbm}
\usepackage{float}
\usepackage{color}
\usepackage{graphicx,subfigure}
\usepackage{multirow,multicol} 
\usepackage{verbatim}
\usepackage{psfrag}
\usepackage{comment}
\usepackage{makeidx}
\usepackage[]{algorithm}
\usepackage{algorithmicx}
\usepackage{algpseudocode}
\usepackage{cases}
 \usepackage{setspace}
\usepackage{pgfplots} 

\newcommand{\mr}[1]{\mathrm{#1}}
\newcommand{\tnr}[1]{{\textnormal{#1}}}

\newcommand{\mc}[1]{\mathcal{#1}}
\newcommand{\mf}[1]{\mathsf{#1}}
 
\newcommand{\ms}[1]{\mathds{#1}}

\newcommand{\ov}[1]{\overline{#1}}

\newcommand{\bh}{\boldsymbol{h}}

\newcommand{\bu}{\boldsymbol{u}}

\newcommand{\bv}{\boldsymbol{v}}
\newcommand{\bw}{\boldsymbol{w}}

\newcommand{\bx}{\boldsymbol{x}}

\newcommand{\bzero}{\boldsymbol{0}}
\newcommand{\bone}{\boldsymbol{1}}

\newcommand{\btheta}{\boldsymbol{\theta}}
\newcommand{\bkappa}{\boldsymbol{\kappa}}

\newcommand{\bxi}{\boldsymbol{\xi}}

\newcommand{\ie}{i.e.,~} 		
\newcommand{\eg}{e.g.,~}	

\newcommand{\argmax}{\mathop{\mr{argmax}}}

\newcommand{\set}[1]{\{#1\}}

\newcommand{\cd}{\cdot}
\newcommand{\ld}{\ldots}

\newcommand{\pdf}{f}            			

\newcommand{\Ex}{\ms{E}}     			
\newcommand{\T}{^{\mf{T}}}            		
\newcommand{\dd}{\,\mr{d}}             		

\newcommand{\mcN}{\mc{N}}

\newcommand{\mfm}{\mf{m}}

\newcommand{\mfP}{\mf{P}}

\newcommand{\Real}{\mathbb{R}}		

\newcommand{\matI}{\tnr{\textbf{I}}}

\newcommand{\matV}{\tnr{\textbf{V}}}

\pgfplotsset{compat=1.12}
\usetikzlibrary{positioning,shadows,shapes,fit,arrows,backgrounds}

\tikzstyle{rect_my} = [draw, rectangle, minimum width=2cm, text width=1.8cm, fill=gray!15, 
  text centered,  minimum height=.9cm]
\tikzstyle{square_my} = [draw, rectangle, minimum width=1cm, text width=0.8cm, fill=gray!15, 
  text centered,  minimum height=.9cm]
\tikzstyle{square_my_graph} = [draw, rectangle, minimum width=1.2cm, text width=1cm, fill=gray!15, 
  text centered,  minimum height=1.2cm]
\tikzstyle{circle_my} = [draw, circle, minimum width=1cm, text width=0.8cm, fill=gray!15, 
  text centered,  minimum height=.9cm]
\tikzstyle{circle_my_graph} = [draw, circle, minimum width=1.1cm, text width=.8cm, fill=gray!15, 
  text centered]
\tikzstyle{cloud_my} = [draw, shape=cloud, minimum width=1cm, text width=0.8cm, fill=gray!15, 
  text centered,  minimum height=1cm]

\tikzstyle{point_my} = [draw=none, minimum width=0cm, text width=0cm, fill=none, 
  text centered,  minimum height=0cm]    
\tikzstyle{line_my} = [draw, -latex]    
\tikzstyle{box_my}=[draw, minimum size=2em, text width=4.5em, text centered]
\tikzstyle{bigbox_my}=[draw, inner sep=15pt]
\tikzstyle{arrow_my} = [thick,->,>=stealth]
\tikzstyle{noarrow_my} = [thick,-,=>stealth]

\definecolor{cblue}{HTML}{1965B0}

\definecolor{cred}{HTML}{B8221E}

\usepackage[acronym,nonumberlist]{glossaries} 
\usepackage{hyperref}
\usepackage{empheq}
\usepackage{verbatim}
\usepackage{graphicx}
\usepackage{enumitem}
\usepackage[title]{appendix}  %
\newacronym[\glsshortpluralkey=PDFs,\glslongpluralkey=probability density functions]{pdf}{PDF}{probability density function}
\newacronym[\glsshortpluralkey=CDFs,\glslongpluralkey=cumulative density functions]{cdf}{CDF}{cumulative density function}
\newacronym[\glsshortpluralkey=CCDFs,\glslongpluralkey=complementary cumulative density functions]{ccdf}{CDF}{complementary cumulative density function}
\newacronym[\glsshortpluralkey=PMFs,\glslongpluralkey=probability mass functions]{pmf}{PMF}{probability mass function}
\newacronym[]{lhs}{l.h.s.}{left-hand side}
\newacronym[]{rhs}{r.h.s.}{right-hand side} 

\newacronym[]{bicm}{BICM}{bit-interleaved coded modulation}
\newacronym[]{bicmid}{BICM-ID}{BICM with iterative demapping}
\newacronym[]{cm}{CM}{coded modulation}
\newacronym[]{tcm}{TCM}{trellis-coded modulation}
\newacronym[]{mlc}{MLC}{multi-level coding}
\newacronym[]{pam}{PAM}{pulse amplitude modulation}
\newacronym[]{bpsk}{BPSK}{binary phase shift keying}
\newacronym[]{qam}{QAM}{quadrature amplitude modulation}
\newacronym[]{16qam}{16-QAM}{16-points quadrature amplitude modulation}
\newacronym[]{psk}{PSK}{phase shift keying}
\newacronym[\glsshortpluralkey=LLRs,\glslongpluralkey=logarithmic likelihood ratios]{llr}{LLR}{logarithmic likelihood ratio}
\newacronym[]{oc}{OC}{operating characteristic}

\newacronym[\glsshortpluralkey=MIs,\glslongpluralkey=mutual informations]{mi}{MI}{mutual information}
\newacronym[\glsshortpluralkey=GMIs,\glslongpluralkey=generalized mutual informations]{gmi}{GMI}{generalized mutual information}
\newacronym[]{eesm}{EESM}{exponential effective-SNR-mapping}
\newacronym[]{bicm-gmi}{BICM-GMI}{BICM generalized mutual information}
\newacronym[]{awgn}{AWGN}{additive white Gaussian noise}
\newacronym[]{bsc}{BSC}{binary symetric channel}
\newacronym[]{amc}{AMC}{adaptive modulation and coding}
\newacronym[]{csi}{CSI}{channel state information}
\newacronym[]{cqi}{CQI}{channel quality indicator}
\newacronym[]{kl}{KL}{Kullback-Leibler}
\newacronym[]{cmm}{CMM}{circular moment matching}
\newacronym[]{ga}{GA}{Gaussian approximation}

\newacronym[]{sp}{SP}{set-partitioning}
\newacronym[]{gsm}{GSM}{global system for mobile communications}
\newacronym[]{edge}{EDGE}{enhanced data rates for GSM evolution}
\newacronym[]{3gpp}{3GPP}{3rd generation partnership project}
\newacronym[]{umts}{UMTS}{Universal Mobile Telecommunication System}
\newacronym[]{lte}{LTE}{Long Term Evolution}
\newacronym[]{dvb}{DVB}{digital video broadcasting}
\newacronym[]{fdd}{FDD}{Frequency Division Duplexing}

\newacronym[\glsshortpluralkey=CCs,\glslongpluralkey=convolutional codes]{cc}{CC}{convolutional code}
\newacronym[\glsshortpluralkey=PCCCs,\glslongpluralkey=parallel concatenated convolutional codes]{pccc}{PCCC}{parallel concatenated convolutional code}
\newacronym[\glsshortpluralkey=TCs,\glslongpluralkey=turbo codes]{tc}{TC}{turbo code}
\newacronym{ldpc}{LDPC}{low-density parity-check}
\newacronym[]{ofdm}{OFDM}{orthogonal frequency-division multiplexing}
\newacronym[]{bep}{BEP}{bit-error probability}
\newacronym[]{wep}{WEP}{word-error probability}
\newacronym[]{sep}{SEP}{symbol-error probability}
\newacronym[]{pep}{PEP}{pairwise-error probability}
\newacronym[]{ttcm}{TTCM}{turbo-trellis coded modulation}
\newacronym[]{uep}{UEP}{unequal error protection}
\newacronym[\glsshortpluralkey=CENCs,\glslongpluralkey=convolutional encoders]{cenc}{CENC}{convolutional encoder}
\newacronym[]{mimo}{MIMO}{multiple-input multiple-output}
\newacronym[\glsshortpluralkey=SNRs,\glslongpluralkey=signal-to-noise ratios]{snr}{SNR}{signal-to-noise ratio}
\newacronym[\glsshortpluralkey=SINRs,\glslongpluralkey=signal-to-interference-plus-noise ratios]{sinr}{SINR}{signal-to-interference-plus-noise ratio}
\newacronym[]{msb}{MSB}{most-significative bit}
\newacronym[]{bcjr}{BCJR}{Bahl--Cocke--Jelinek--Raviv}
\newacronym[]{cbc}{CBC}{Colavolpe--Barbieri--Caire}
\newacronym[]{skr}{SKR}{Shayovitz--Kreimer--Raphaeli}
\newacronym[\glsshortpluralkey=SEDs,\glslongpluralkey=squared Euclidean distances]{sed}{SED}{squared Euclidean distance}
\newacronym[\glsshortpluralkey=EDs,\glslongpluralkey=Euclidean distances]{ed}{ED}{Euclidean distance}
\newacronym[\glsshortpluralkey=MEDs,\glslongpluralkey=minimum Euclidean distances]{med}{MED}{minimum Euclidean distance}
\newacronym[]{core}{CoRe}{constellation rearrangement}
\newacronym[]{pdl}{PDL}{parallel decoding of the individual levels}
\newacronym[\glsshortpluralkey=GCs,\glslongpluralkey=Gray codes]{gc}{GC}{Gray code}
\newacronym[]{brgc}{BRGC}{binary-reflected Gray code}
\newacronym[]{nbc}{NBC}{natural binary code}
\newacronym[]{fbc}{FBC}{folded-binary code}
\newacronym[]{bsgc}{BSGC}{binary semi-Gray code}
\newacronym[]{msp}{MSP}{modified set-partitioning}
\newacronym[]{ssp}{SSP}{semi set-partitioning}
\newacronym[]{fhd}{FHD}{free Hamming distance}
\newacronym[]{mfhd}{MFHD}{maximum free Hamming distance}
\newacronym[]{ods}{ODS}{optimal distance spectrum}
\newacronym[]{iud}{i.u.d.}{independent and uniformly distributed}
\newacronym[]{ud}{u.d.}{uniformly distributed}
\newacronym[]{iid}{i.i.d.}{independent, identically distributed}
\newacronym[]{ami}{AMI}{accumulated mutual information}
\newacronym[]{bico}{BICO}{binary-input continuous-output}
\newacronym[]{gh}{GH}{Gauss--Hermite}
\newacronym[]{gum}{GUM}{Gaussian--uniform mixture}

\newacronym[\glsshortpluralkey=BSs,\glslongpluralkey=base-stations]{bs}{BS}{base-station}
\newacronym[\glsshortpluralkey=MSs,\glslongpluralkey=mobile-stations]{ms}{MS}{mobile-stations}

\newacronym[]{phy}{PHY}{physical layer} 
\newacronym[]{rlc}{RLC}{Radio-Link control} 
\newacronym[]{ran}{RAN}{Radio Access Network} 
\newacronym[]{llc}{LLC}{logical link control} 
\newacronym[]{tcp}{TCP}{transmission control protocol} 
\newacronym[]{mac}{MAC}{media access control} 
\newacronym[]{fft}{FFT}{fast Fourier transform} 
\newacronym[]{ft}{FT}{Fourrier transform}
\newacronym[]{cf}{CF}{characteristic function} 
\newacronym[]{mgf}{MGF}{moment generating function} 
\newacronym[]{ee}{EE}{energy efficiency} 
\newacronym[]{eb}{EB}{energy per bit}
\newacronym[]{kkt}{KKT}{Karush--Kuhn--Tucker} 
\newacronym[]{mcs}{MCS}{modulation/coding scheme} 
\newacronym[]{fec}{FEC}{forward error correction}
\newacronym[]{arq}{ARQ}{automatic repeat request}
\newacronym[]{harq}{HARQ}{hybrid ARQ}
\newacronym[]{tarq}{TARQ}{truncated HARQ}
\newacronym[]{ir}{IR}{incremental redundancy}
\newacronym[]{rpr}{RR}{repetition redundancy}
\newacronym[]{rrharq}{RR-HARQ}{repetition redundancy HARQ}
\newacronym[]{irharq}{IR-HARQ}{incremental redundancy HARQ}
\newacronym[]{ack}{ACK}{positive acknowledgment}
\newacronym[]{nack}{NACK}{negative acknowledgment}
\newacronym[]{hol}{HoL}{head of the line}
\newacronym[]{crc}{CRC}{cyclic redundancy check}
\newacronym[]{dp}{DP}{dynamic programming}
\newacronym[]{gp}{GP}{geometric programming}
\newacronym[]{per}{PER}{packet error rate}
\newacronym[]{ber}{BER}{bit error rate}
\newacronym[]{op}{OP}{outage probability}
\newacronym[]{spa}{SPA}{saddle-point approximation}
\newacronym[]{mrc}{MRC}{maximum ratio combining}
\newacronym[]{mdp}{MDP}{Markov decision process}
\newacronym[]{lp}{LP}{linear programming}
\newacronym[]{pomdp}{POMDP}{partially observable Markov decision process}
\newacronym[]{psimdp}{PSI-MDP}{partial state information Markov decision process}
\newacronym[]{scpp}{SCPP}{stochastic shortest path problem}

\newacronym[]{forw}{frwd}{forward}
\newacronym[]{feed}{fdbk}{feedback}

\newacronym[]{mm}{MM-HARQ}{multi-message HARQ}
\newacronym[]{xp}{XP-HARQ}{cross-packet HARQ}
\newacronym[]{ts}{TS}{time-sharing}
\newacronym[]{sc}{SC}{superposition coding}
\newacronym[]{sbrq}{SBRQ}{systematic backward retransmission}
\newacronym[]{brq}{BRQ}{backward retransmission}
\newacronym[]{lharq}{L-HARQ}{layer-coded HARQ}
\newacronym[]{anlharq}{AoN-HARQ}{all-or-none L-HARQ}
\newacronym[]{vlharq}{VL-HARQ}{variable-length HARQ}

\newacronym[]{pp}{PPP}{point process}
\newacronym[]{ppp}{PPP}{Poisson point process}

\newacronym[]{fide}{FIDE}{F\'ed\'eration Internationale des \'Echecs}
\newacronym[]{fifa}{FIFA}{F\'ed\'eration Internationale de Football Association}
\newacronym[]{fivb}{FIVB}{F\'ed\'eration Internationale de Volleyball}
\newacronym[]{epl}{EPL}{English Premier League}
\newacronym[]{nhl}{NHL}{National Hockey League}
\newacronym[]{shl}{SHL}{Swedish Hockey League}
\newacronym[]{nfl}{NFL}{National Football League}
\newacronym[]{ipl}{IPL}{Indian Premier League}
\newacronym[]{nba}{NBA}{National Basketball Association}
\newacronym[]{mls}{MLS}{Major League Soccer}

\newacronym[]{sg}{SG}{stochastic gradient}
\newacronym[]{lms}{LMS}{least mean squares}
\newacronym[]{nlms}{NLMS}{normalized LMS}
\newacronym[]{rls}{RLS}{recursive least squares}
\newacronym[]{vss}{VSS}{variable step-size}
\newacronym[]{hfa}{HFA}{home-field advantage}
\newacronym[]{ha}{HA}{home advantage}
\newacronym[]{mov}{MOV}{margin of victory}
\newacronym[]{ac}{AC}{adjacent categories}
\newacronym[]{cl}{CL}{cumulative link}
\newacronym[]{glm}{GLM}{generalized linear models}
\newacronym[]{nn}{NN}{neural networks}
\newacronym[]{rps}{RPS}{ranked probability score}
\newacronym[]{mse}{MSE}{mean square error}
\newacronym[]{mmse}{MMSE}{minimum mean square error}
\newacronym[]{rmse}{RMSE}{root mean squared error}
\newacronym[]{ols}{OLS}{ordinary least squares}
\newacronym[]{map}{MAP}{maximum a posteriori}
\newacronym[]{ml}{ML}{maximum likelihood}
\newacronym[]{loo}{LOO}{leave-one-out}
\newacronym[]{alo}{ALO}{approximate leave-one-out}
\newacronym[]{logo}{LOGO}{leave-one-game-out}
\newacronym[]{alogo}{ALOGO}{approximate leave-one-game-out}
\newacronym[]{msd}{MSD}{mean-square deviation}
\newacronym[]{lop}{LOP}{linear ordering problem}
\newacronym[]{so}{SO}{shootouts}
\newacronym[]{rt}{RT}{regulation time}
\newacronym[]{ot}{OT}{overtime}
\newacronym[]{rr}{RR}{round-robin}
\newacronym[]{irt}{IRT}{item-response theory}

\newacronym[]{dmp}{DMP}{discretized message passing}
\newacronym[]{mp}{MP}{message passing}
\newacronym[]{ep}{EP}{expectation propagation}
\newacronym[]{em}{EM}{expectation maximization}
\newacronym[]{hmm}{HMM}{hiden Markov models}

\newacronym[]{svd}{SVD}{singular values decomposition}

\newacronym[]{vkf}{vKF}{\emph{vector-variance} Kalman Filter}
\newacronym[]{skf}{sKF}{\emph{scalar-variance} Kalman Filter}
\newacronym[]{fkf}{fKF}{\emph{fixed-variance} Kalman Filter}
\newacronym[]{kf}{KF}{Kalman filter}
\newacronym[]{gelo}{G-Elo}{generalized Elo}
\newacronym[]{mvdr}{MVDR}{minimum variance distortionless response}
\newacronym[]{lcmv}{LCMV}{linearly-constrained minimum variance}
\newacronym[]{music}{MUSIC}{multiple signal classification}
\newacronym[]{cp}{CP}{canonical polyadic}

\newacronym[]{tpb}{TPB}{tensor-product-basis}

\newtheorem{lemma}{Lemma}

\newtheorem{example}{Example}

\begin{document}

\title{A Unified Bayesian Perspective for\\
Conventional and Robust Adaptive Filters}

\author[1]{Leszek~Szczecinski\corref{cor1}}
\ead{leszek.szczecinski@inrs.ca}

\author[1]{Jacob~Benesty}
\ead{jacob.benesty@inrs.ca}

\author[2]{Eduardo~Vinicius~Kuhn}
\ead{kuhn@utfpr.edu.br}

\cortext[cor1]{Corresponding author.}

\address[1]{INRS–Institut National de la Recherche  Scientific, Montreal, QC, H5A-1K6, Canada.}
\address[2]{LAPSE–Electronics and Signal Processing Laboratory, Department of Electronics Engineering, Federal University of Technology - Paran\'a, Toledo, Paran\'a, 85902-490, Brazil.}

\begin{abstract}
    In this work, we present a new perspective on the origin and interpretation of adaptive filters. By applying Bayesian principles of recursive inference from the state-space model and using a series of simplifications regarding the structure of the solution, we can present, in a unified framework, derivations of many adaptive filters that depend on the probabilistic model of the measurement noise. In particular, under a Gaussian model, we obtain solutions well-known in the literature (such as LMS, NLMS, or Kalman filter), while using non-Gaussian noise, we derive new adaptive algorithms. Notably, under the assumption of Laplacian noise, we obtain a family of robust filters of which the sign-error algorithm is a well-known member, while other algorithms, derived effortlessly in the proposed framework, are entirely new. Numerical examples are shown to illustrate the properties and provide a better insight into the performance of the derived adaptive filters. 
\end{abstract}


\maketitle

\section{Introduction}\label{Sec:Introduction}

In this work, we propose a Bayesian perspective that can be used to explain the origins and properties of conventional and robust adaptive filters, which are arguably among the most important tools in the area of signal processing with countless applications such as adaptive control \cite{astrom2008}, system identification, noise and echo cancellation \cite{benesty2001}, active noise cancellation \cite{kuo1996active}, channel equalization \cite[Ch.~5.4]{Sayed08_Book}, antenna processing, adaptive beamforming \cite[Ch.~6.5]{Sayed08_Book}, \cite{chandran2004adaptive}, and many others \cite{diniz2013,farhang2013,haykin2014}.

The Bayesian formulation was often used in the estimation literature \eg in \cite{Roweis99,Zayyani09,Szczecinski21,Lange24}, and is also not new in signal processing applications. For example, Bayesian principles can be used to derive the well-known Kalman filter \cite{Roweis99}, \cite{Sadiki06}, and \cite[Ch.~13.2]{Moon00_Book}, or to improve the performance of adaptive filters \cite{Deng05,Huemmer15}. On the other hand, we are not aware of a Bayesian framework in which the most popular adaptive filters are derived from the common model.

Our work aims at filling this gap: By using the well-known state-space model, we show how the adaptive algorithms with a varying degree of complexity can be derived in a systematic way. In particular, for a Gaussian state-space model, the well-known adaptive filters such as least-mean-square (LMS) \cite{widrow1960,widrow1975}, normalized LMS (NLMS) \cite{kaczmarz1937,nagumo1967}, and others can be derived as particular cases of \gls{kf} \cite{kalman1960new,kalman1961new}. Such a perspective not only has a unifying and educational value, allowing us to see all adaptive algorithms in a common framework, but it also leads to a simple and meaningful interpretation of their parameters (such as adaptation step and/or regularization coefficient) in terms of the statistical quantities (such as variances of the estimates).

More importantly, the Bayesian perspective we propose allows us to derive new adaptive algorithms suited to deal with non-Gaussian models. In particular, we are interested in robust filters to deal with frequent outliers. Indeed, using the generalized Gaussian distribution of the measurement noise in the state-space model, we derive an entire class of new robust adaptive algorithms: by specializing the derived algorithms to Laplacian noise (which is a particular case of a generalized Gaussian distribution), we derive algorithms which are natural generalizations of the sign-error LMS algorithm \cite{hirsch1970simple,claasen1981comparison,gersho1984adaptive,mathews1987improved}, which is a well-known adaptive filter commonly used to deal well with outliers.

In the non-Gaussian formulation, the problems we deal with have no closed-form solutions, and thus the filtering equations are approximate, where the closed-form solutions can be enhanced by iterative operations. This is an intriguing and rarely used approach in signal processing, which, as we show, can improve the performance of adaptive filtering.

The algorithms are tested using synthetic data with numerical results validating the derivations, where the performance deteriorates when we increase the number of simplifying assumptions. The performance is notably improved by the new robust adaptive filters in the presence of non-Gaussian noise.

\section{Approximate Bayesian inference in state-space models}\label{Sec:Bayesian.inference}

Consider the following state-space model:
\begin{align}
\label{state.equation.general}
        \btheta_{t} &=\btheta_{t-1} + \bxi_t,\\
\label{output.equation.general}
    y_t &= \bx_t\T\btheta_t + \eta_t ,
\end{align}
where $\btheta_t\in\Real^M$ is the random vector-state at time $t=1,2,\ld$, $y_t\in\Real$ -- the observation at time $t$ which depends on the linear function of the state $\bx_t\T\btheta_t$, where $\bx_t\in\Real^M$ is known and $(\cd)\T$ is the transpose operator, $\eta_t$ is the measurement/observation noise modeled as a zero-mean, random variable with distribution $\pdf(\eta_t)$, and $\bxi_t\in\Real^M$, $t=1,2,\ld$ are independent zero-mean Gaussian vectors composed of independent elements: 
\begin{align}
\label{pdf(u_t)}
    \pdf(\bxi_t) &= \mcN(\bxi_t; \bzero, \varepsilon_t \matI),
\end{align}
where $\mcN(\bu;\bw,\matV)$ is the Normal \gls{pdf} in $\bu$ with the mean $\bw$ and covariance matrix $\matV$. For simplicity of derivations, we use $\varepsilon_t=\varepsilon$. Moreover, we assume that the unconditional prior distribution of the state $\btheta_0$ is given by $\pdf(\btheta_0)$. 

The state dynamics defined in \eqref{state.equation.general} is a simplified version of a more general linear model, \eg \cite[Sec.~2]{Roweis99}, where the covariance of $\bxi_t$ does not need to be diagonal. However, it is sufficient to derive known adaptive filters from the general estimation framework because we rely on simplified models with diagonal covariances. Note also that we do not use \eqref{state.equation.general} to model a particular/known relationship between the element of $\btheta_t$. Rather, it is introduced to allow the algorithms to track the time-varying parameters $\btheta_t$ and/or control the convergence even if $\btheta_t=\btheta$, \ie when the state does not change in time, as we will see in the numerical examples. 

In the Bayesian formulation, the estimation consists in finding the posterior distribution of $\btheta_t$ from all available information at time $t$, \ie $y_{1:t}=\set{y_1,\ld, y_t}$, and it  can be calculated as \cite[Ch.~12]{Moon00_Book}
\begin{align}
    \pdf(\btheta_t| y_{1:t})
    &=
\pdf(\btheta_t|  y_{1:t-1}, y_t )
=
\frac{\pdf(y_t, \btheta_t, y_{1:t-1})}{\pdf(y_{1:t-1}, y_t )}\\
&\propto
\pdf(y_t|\btheta_t) \pdf(\btheta_t| y_{1:t-1}) ,
\end{align}
where  
\begin{align}
\pdf(\btheta_t| y_{1:t-1})
\label{theta.aposteriori.2}
&=
\int  
\pdf(\btheta_t|\btheta_{t-1})
\pdf(\btheta_{t-1}|  y_{1:t-1})
\dd \btheta_{t-1};
\end{align}
we also use $f(\btheta)\propto \phi(\btheta)$ to indicate that a normalization may be required to obtain a distribution, \ie $ f(\btheta) = \phi(\btheta)/\int \phi(\btheta) \dd \btheta$.

From \eqref{state.equation.general} and \eqref{pdf(u_t)}, we easily obtain
\begin{align}
\label{pdf.theta.t.theta.t1}
    \pdf(\btheta_t | \btheta_{t-1}) = \mcN(\btheta_t;\btheta_{t-1},\varepsilon \matI),
\end{align}
and, from \eqref{output.equation.general}, we know that
\begin{align}
\label{pdf.y_t.generalized.linear}
    \pdf(y_t|\btheta_t) &\equiv \pdf_{\eta_t}(y_t-\bx_t\T\btheta_t).
\end{align}

In particular, if $\eta_t$ is a zero-mean Gaussian noise with variance $v_\eta$, \ie $\pdf(\eta_t) =\mcN(\eta_t;0,v_{\eta})$, \eqref{pdf.y_t.generalized.linear} is given by
\begin{align}
\label{y_t.Gaussian}
    \pdf(y_t|\btheta_t) &= \mcN(y_t-\bx_t\T\btheta_t; 0, v_\eta),
\end{align}
and if $\pdf(\btheta_0)$ is also Gaussian, all variables in the model are Gaussian; so, \eqref{theta.aposteriori.2} can be obtained in closed-form, yielding the Kalman filter equations \cite[Sec.~5.4]{Roweis99}. 

In general, however, a closed-form solution to \eqref{theta.aposteriori.2} is not available; hence, we may need approximations or explicit numerical methods. The latter are well-known in the Bayesian estimation literature and calculate the integral \eqref{theta.aposteriori.2} via Monte Carlo methods; this produces the so-called particle filtering algorithms \cite{Talebi17}. On the other hand, in the signal processing context, the adaptive filters require simple implementation and thus the complexity of the particle filtering is not acceptable. The approximations are then more suitable \cite{Talebi23,Lange24} and we adopt this strategy in our work not only due to its simplicity but also because this is how well-known adaptive algorithms can be derived in a common framework.

Note that we abuse the notation and the distributions $\pdf(\cd)$ and $\pdf(\cd|\cd)$ are only identified by their arguments. When it might lead to confusion, we are specific, \eg as in \eqref{pdf.y_t.generalized.linear}, where $\pdf_{\eta_t}(e)$ denotes the \gls{pdf} of $\eta_t$ evaluated for $\eta_t=e$.

\subsection{Gaussian approximations}\label{Sec:Gaussian.approximation}

The simplest approach is to approximate the terms in \eqref{theta.aposteriori.2} using Gaussian distributions (denoted by tilde), \ie
\begin{align}
\label{pdf.theta.t.approx.as.Gaussian}
    \pdf(\btheta_t|y_{1:t})&\approx\tilde\pdf(\btheta_t|y_{1:t})=\mcN(\btheta_t;\bw_t,\matV_t),
\end{align}
thereby, we can easily calculate \eqref{pdf.theta.t.theta.t1} as follows:
\begin{align}
\label{integral.theta_t_1}
    \tilde\pdf(\btheta_t|y_{1:t-1})
    &=\int  \pdf(\btheta_{t}|\btheta_{t-1})\tilde{\pdf}(\btheta_{t-1}|  y_{1:t-1}) \dd \btheta_{t-1}\\
    &=\int \mcN(\btheta_t;\btheta_{t-1},\epsilon\matI)\mcN(\btheta_{t-1},\bw_{t-1},\matV_{t-1})\dd\btheta_{t-1}\\
\label{theta.t_1.is.Gaussian}
    &=
\mcN(  \btheta_{t} ; \bw_{t-1} , \ov{\matV}_t  ),
\end{align}
where we used \eqref{pdf.theta.t.theta.t1}, \eqref{pdf.theta.t.approx.as.Gaussian}, and 
\begin{align}
\label{ov.matV}
    \ov{\matV}_t = \matV_{t-1} +\varepsilon \matI
\end{align}
is the covariance matrix of the $\btheta_t$ conditioned on $y_{1:t-1}$ (while $\matV_{t-1}$ is the covariance of $\btheta_{t-1}$ conditioned on the same observations).

Then, we can use \eqref{theta.t_1.is.Gaussian} to approximate  \eqref{theta.aposteriori.2} as follows:
\begin{align}
\label{Projection}
    \tilde \pdf(\btheta_{t}|  y_{1:t}) 
&=
\mfP\big[\phi(\btheta_t|y_{1:t}) \big]=\mcN(\btheta_t; \bw_t, \matV_t),\\
\label{phi(theta_t)}
\phi(\btheta_t|y_{1:t})
&=\pdf(y_t|\btheta_t)
\tilde\pdf(\btheta_{t}|y_{1:t-1})
=\pdf(y_t|\btheta_t)\mcN(  \btheta_{t} ; \bw_{t-1} , \ov{\matV}_t  ),
\end{align}
where $\mfP\big[\phi(\btheta_t|y_{1:t})\big]$ is an operator ``projecting" $\phi(\btheta_t|y_{1:t})$ into the space of Gaussian distributions, which boils down to finding the mean $\bw_t$ and covariance $\matV_t$ of the distribution $\phi(\btheta_t|y_{1:t})$. Note that $\phi(\btheta_t|y_{1:t})$ (the distribution before projection) already contains approximations due to the use of $\tilde\pdf(\btheta_t|y_{1:t-1})$. However, the projection is necessary as the product $\phi(\btheta_t|y_{1:t})$ is not Gaussian in $\btheta_t$ (because, in general, $\log\pdf(y_t|\btheta_t)$ is not a quadratic function of $\btheta_t$).

To define the projection operator $\mfP[\cd]$, we might find the Gaussian \gls{pdf} which minimizes the \gls{kl} divergence with $\phi(\btheta_t|y_{1:t})$, in which case we obtain
\begin{align}
\label{KL.mean}
    \bw_t & = \Ex[\btheta_t] , \\
\label{KL.variance}
    \matV_t &= \Ex[\btheta_t\btheta_t\T]-  \bw_t\bw_t\T,
\end{align}
where the expectation is calculated with respect to the distribution $\phi(\btheta_t|y_{1:t})$ being projected. The latter requires integration (over the $M$-dimensional space of $\btheta_t$), which is impractical. 

Thus, in search for simpler solutions, we approximate the mean with the mode of the distribution and the variance with the inverted Hessian of its logarithm, \ie
\begin{align}
\label{mu_t.from.max}
    \bw_t & = \argmax_{\btheta_t} \log \phi(\btheta_t|y_{1:t}) , \\
\label{V_t.from.Hessian}
    \matV_t & = [-\nabla_{\btheta_t}^2 \log \phi(\btheta_t|y_{1:t})\big|_{\btheta_t=\bw_t}]^{-1}.
\end{align}
Only if  $\phi(\btheta_t|y_{1:t})$ is Gaussian, \eqref{mu_t.from.max}-\eqref{V_t.from.Hessian} are equivalent to \eqref{KL.mean}-\eqref{KL.variance}.

\subsection{Approximate Kalman filter}\label{Sec:Approximate.Kalman}

The projection \eqref{Projection} requires solving the problem \eqref{mu_t.from.max}:
\begin{align}
\label{w_t=argmax}
    \bw_t &= \argmax_{\btheta}
    \Big\{ \ell(y_t-\bx_t\T\btheta) +\log \tilde\pdf_{\btheta_t}(\btheta|y_{1:t-1})\Big\} ,
\end{align}
where
\begin{align}
\label{ell.definition}
    \ell(e)&=\log \pdf_{\eta_t}(e).
\end{align}

Since $\tilde\pdf(\btheta_t|y_{1:t-1})$ is Gaussian, $\log\tilde\pdf(\btheta_t|y_{1:t-1})$ is quadratic in $\btheta_t$, so \eqref{w_t=argmax} can be solved in closed form if we approximate $\ell(e)$ by a concave and quadratic function of $e$, \ie if instead of \eqref{w_t=argmax}, we solve its approximate version:
\begin{align}
\label{w_t=argmax.approx}
    \bw_t &\approx \argmax_{\btheta_t}
    \Big\{q(y_t-\bx_t\T\btheta_t) +\log \tilde\pdf(\btheta_t|y_{1:t-1})\Big\} ,
\end{align}
where
\begin{align}
\label{ell(z).approx.quadratic}
    \ell(e)\approx q(e)= - \frac{1}{2}h_t e^2 + C_q ,
\end{align}
with $C_q$ being a constant, independent of $e$, 
and 
\begin{align}
    h_t = h(e_t) >0,
\end{align}
\ie the form of the approximation may depend on the approximation error $e_t$. Only in the Gaussian case we have a constant function $h(e)=\frac{1}{v_\eta}$.

Note that we limit our attention to the case $q(e)=q(-e)$ because in most practically interesting cases the noise $\eta_t$ has a symmetric distribution, i.e., $\pdf_{\eta_t}(e)=\pdf_{\eta_t}(-e)$.

The approximation sign $\approx$ will be dropped in the following, and we will use it only when a new approximation is introduced.

Now, using \eqref{ell(z).approx.quadratic} and \eqref{theta.t_1.is.Gaussian} in \eqref{w_t=argmax.approx}, after simple algebra yields
\begin{align}
\label{ov.matV_t.Kalman.full}
    \ov{\matV}_t &= \matV_{t-1} +\varepsilon \matI,\\
\label{partial.Kalman.gain.full}
    \bkappa_t&=\ov{\matV}_t\bx_t,\\
\label{omega_t.definition}
    s_t &= \bx_t\T\bkappa_t,\\
\label{e_t.definition}
    e_t&= y_t-\bx_t\T\bw_{t-1},    \\
\label{h.t.definition}
    h_t & = h(e_t),\\
\label{a_t.definition}
   \alpha_t & = \frac{h_t}{1+h_t s_t},\\
\label{w_t.update}
    \bw_{t}&=\bw_{t-1} + \bkappa_t\alpha_t e_t,\\
\label{V_t.update}
    \matV_{t}
    &=
    \ov{\matV}_t(\matI -\bx_{t}\bkappa_t\T\alpha_t),
\end{align}
where $e_t$ is the prediction error, $s_t$
is the second moment of the predictor $\bx_t\T\btheta_t$ calculated from the conditional distribution $\pdf(\btheta_t|y_{1:t-1})$, and the update of the covariance matrix in \eqref{V_t.update} is obtained from \eqref{V_t.from.Hessian} via matrix inversion lemma.

In the Gaussian case, \ie if $h_t=\frac{1}{v_\eta}$, the operations in \eqref{ov.matV_t.Kalman.full}--\eqref{V_t.update} become the well-known \gls{kf} estimation algorithm for the model \eqref{state.equation.general}--\eqref{output.equation.general}.
Here $\bkappa_t$ is the ``preliminary" Kalman gain vector, and $\alpha_t$ is the gain multiplier (so the actual Kalman gain is given by the product $\bkappa_t\alpha_t$). This separation will be conceptually useful in non-Gaussian case, which we treat in the following.

To initialize the recursion \eqref{ov.matV_t.Kalman.full}-\eqref{V_t.update}, we can use, \eg $\bw_0=\bzero$ and $\matV_0=v_0\matI$, where $v_0$ -- the prior variance of elements of $\btheta_0$ -- is a parameter of the algorithm which we must define.

\subsection{Differences with the conventional Kalman filter}
The presentation of the problem \eqref{w_t=argmax} points to two issues that do not appear in the conventional Kalman filter derived from the assumption that the noise $\eta_t$ is Gaussian.

First, we note that the quadratic approximation of $\ell(e_t)$ may be defined in many ways. Here we postulate to choose it so as to guarantee that the update step \eqref{w_t.update}, which solves \eqref{w_t=argmax.approx}, also increases the objective  function in \eqref{w_t=argmax}, \ie we want to guarantee that
\begin{align}
\nonumber
    \big[\ell(y_t-\bx_t\T\btheta_t)
    &+\log\tilde\pdf(\btheta_t|y_{1:t-1})\big]_{\btheta_t=\bw_t}\\
\label{increase.objecfive.condition}
    &>
    \big[\ell(y_t-\bx_t\T\btheta_t)+\log\tilde\pdf(\btheta_t|y_{1:t-1})\big]_{\btheta_t=\bw_{t-1}}.
\end{align}
This requirement is quite natural, as we expect the approximation not only to provide us with closed-form expressions but also to address meaningfully the maximization problem \eqref{w_t=argmax}, \ie at least we want the objective function under maximization to increase.

It is simple to show that, by applying minorization:
\begin{align}\label{minorization}
    q(e) &\le \ell(e), \quad e\neq e_t , \\    \label{minorization.equality}
    q(e_t)&= \ell(e_t) ,
\end{align}
we guarantee that \eqref{increase.objecfive.condition} is, indeed, satisfied. This approach is inspired by the minorization-maximization \cite{Hunter04a}.

The second issue is that the maximization in \eqref{w_t=argmax} is solved only approximately by \eqref{w_t.update} and may be improved by  invoking \eqref{e_t.definition}-\eqref{w_t.update} iteratively, as follows:
\begin{align}
\label{w_t.0}
\bw_{t,0}&=\bw_{t-1}\\
    \label{w_t.update.iterations}
&\begin{rcases}
    e_{t,i} = y_t - \bx_t\T\bw_{t,i}\\ 
    h_{t,i}=h(e_{t,i})\\
    \alpha_{t,i}=\displaystyle \frac{h_{t,i}}{1+h_{t,i} s_t}\\
    \bw_{t,i+1}=\bw_{t-1} + \bkappa_{t} \alpha_{t,i}e_{t,0}\\
\end{rcases} \quad i=0, \ld, I\\
\alpha_{t}&=\alpha_{t,I}\\
\label{w_t.I+1}
\bw_t&=\bw_{t,I+1},
\end{align}
where $\alpha_{t,i}$  is the gain multiplier updated iteratively.

For $I=0$, operations in \eqref{w_t.update.iterations} are equivalent to \eqref{w_t.update}. In other words, $I$ is the number of iterations executed beyond the one-step update \eqref{w_t.update} and, by the minorization form of $q(e_t)$, we have a guarantee that the function under maximization in \eqref{w_t=argmax} cannot decrease. Note that, if $h(e)$ is constant, $\alpha_{t,i}$ will not be affected by the iterations and thus $\bw_{t,i+1}$ will not change with $i$. Therefore, in the conventional Gaussian case, when $h(e)=\frac{1}{v_\eta}$, there is no point in using this iterative approach.

\subsection{Simplifications for adaptive Filtering}\label{Sec:Simplification.covariance}

We take one more step to come closer to the reality of adaptive filtering. In particular, we restrict the covariance $\matV_t$ produced via the projection $\mfP[\cd]$ in \eqref{Projection} as follows:
\begin{align}\label{V_t.constrained}
    \matV_t=
    \begin{cases}
        \tnr{diag}(\bv_t)&\tnr{vector-variance model}\\
        v_t\matI&\tnr{scalar-variance model}\\
        v\matI&\tnr{fixed-variance model}
    \end{cases},
\end{align}
which corresponds to adopting different models for $\btheta_t$, namely:
\begin{itemize}
    \item in the vector-variance model, we assume that the parameters $\btheta_t$ are independent Gaussian variables, characterized by the means and variances kept, respectively, in $\bw_t$ and in $\bv_t$;

    \item in the scalar-variance model, we assume that the variance of all elements of $\btheta_t$ is equal to $v_t$; while

    \item in the fixed-variance model, the variance is also common to all elements of $\btheta_t$, but it does not change as $t$ evolves. 
\end{itemize}

\begin{lemma}\label{Lemma:1}
    If $\matV_t$ is obtained through \eqref{KL.variance}, the \gls{kl} divergence is minimized for
\begin{align}
\label{vec_t.from.diagonal}
    \bv_t = \tnr{di}(\matV_t),
\end{align}
where  $\tnr{di}(\matV)$ is the diagonal of $\matV$, 
while $v_t$ is obtained as the average of the diagonal elements in $\matV_{t}$, \ie
\begin{align}
\label{v_t.from.avarage}
    v_t = \frac{\tnr{Tr}(\matV_t)}{M}=\frac{\bone\T\bv_t}{M},
\end{align}
where $\tnr{Tr}(\cd)$ is the trace of the matrix.

\textbf{Proof}: see \cite[Appendix~A]{Szczecinski21}.
\end{lemma}

Note that we will use \eqref{vec_t.from.diagonal} and \eqref{v_t.from.avarage} even if the projection is carried out approximately [\ie when we use \eqref{mu_t.from.max}-\eqref{V_t.from.Hessian} instead of the ``exact" \gls{kl} projection \eqref{KL.mean}-\eqref{KL.variance}].

To derive new adaptive algorithms, we replace the covariance $\matV_t$ with a diagonal matrix given by the model \eqref{V_t.constrained}, where, depending on how we constrain the covariance, we will obtain the \gls{vkf} [where we apply \eqref{vec_t.from.diagonal}], the \gls{skf} [where we use \eqref{v_t.from.avarage}], and the \gls{fkf} algorithms [where we always use $\matV= \ov{v}\matI$, see \eqref{V_t.constrained}], which we summarize as follows:
\begin{align}
\nonumber
    &\tnr{vKF}& &\tnr{sKF}&  &\tnr{fKF}&\\
    &\ov\bv_{t}= \bv_{t-1}+\varepsilon \bone&
    &\ov{v}_{t}= v_{t-1}+\varepsilon&
    \\
    &\bkappa_t=\ov\bv_{t}\odot\bx_t&
    &\bkappa_t=\ov{v}_{t}\bx_t&
    &\bkappa_t=\ov{v} \bx_t&
    \\
    &s_t = \bx_t\T\bkappa_t&
    &s_t = \ov{v}_t\|\bx\|^2&
    &s_t = \ov{v}\|\bx\|^2&\\
    \nonumber
    &&&\vdots\\
    \nonumber
    &&&\tnr{Eqs.~}\eqref{e_t.definition}-\eqref{w_t.update}
    \\
    \nonumber
    &&&\tnr{or}\quad \tnr{Eqs.}\eqref{w_t.0}-\eqref{w_t.I+1}&
    \\
    \nonumber
    &&&\vdots\\
\label{variance.update}
    &\bv_{t} = \ov\bv_{t} \odot(\bone  - \bkappa_t\odot \bx_t\alpha_t)&
    &v_{t}=
    \ov{v}_{t} \Big(1 - s_t\alpha_t/M
    \Big),
\end{align}
where $\odot$ denotes the element-by-element multiplication between vectors.



The above equations can be expressed compactly. In particular,  the \gls{fkf} algorithm may be written as
\begin{align}\label{fixed.variance}
    \bw_{t}&=\bw_{t-1} + \bx_t\frac{\ov{v}h_te_t}{1+\ov{v} h_t\|\bx_t\|^2},
\end{align}
that, for large posterior variance,  $\ov{v}$, becomes
\begin{align}\label{fixed.variance.large.v}
    \bw_{t}&=\bw_{t-1} + \bx_t\frac{e_t}{\|\bx_t\|^2},
\end{align}
which is the \gls{nlms} algorithm.

Furthermore, for $h_t \|\bx_t\|^2 \ov{v} \ll 1$, we obtain
\begin{align}
\label{SG.algoritm}
    \bw_{t}
    &=\bw_{t-1} + \ov{v}\bx_t h_t e_t,\\
\label{SG.algoritm.2}
    &=\bw_{t-1} + \ov{v} \Big[\nabla_{\btheta} q(y_t-\bx_t\T\btheta)\Big]_{\btheta=\bw_{t-1}},
\end{align}
which we recognize \eqref{SG.algoritm.2} to be the \gls{sg} maximization (with respect to $\btheta$) of the function $q(y_t-\bx_t\T\btheta)$ with the adaptation step $\ov{v}$, which is equal to the assumed posterior variance of the elements in $\btheta_t$.

\subsection{Sanity check: Gaussian case}\label{Sec:Sanity.check.Gaussian.case}
As a sanity check, before delving into non-Gaussian noise $\eta_t$, let us assume that $\eta_t$ are \gls{iid} zero-mean, Gaussian variables with variance $v_\eta$; so, using  \eqref{y_t.Gaussian}, we have
\begin{align}
    \ell(e) &= q(e)= -\frac{e^2}{2v_\eta}+\tnr{Const.} ,\\
\label{h_t.Gaussian}
    h_t & = \frac{1}{v_{\eta}}.
\end{align}
Thus, the quadratic representation of $\ell(e)$ is unique, independent of $e$ and, therefore, no iterations are needed, \ie we may set $I=0$.

It is easy to see that the algorithms derived in the general case correspond to those already known from the literature, as explained below.
\begin{itemize}
    \item \gls{sg} algorithm, \eqref{SG.algoritm} yields
\begin{align}
\label{LMS.algorithm}
    \bw_{t}&=\bw_{t-1} + \mu\bx_t e_t,\\
    \mu & = \frac{\ov{v}}{v_\eta},
\end{align}
which is the LMS algorithm. Its step, $\mu$, is defined as a ratio between the variance of the estimate and the variance of the measurement noise, and we recognize it as the limit case of \cite[Eq.~(5)]{Huemmer15}. Although the simplicity of this interpretation of the LMS step is appealing, it is a reminder of the series of approximations we adopted because we know from the literature, \eg \cite[Ch.~16]{Sayed08_Book}, that the optimal step depends on the eigenvalues of the covariance matrix of the data. 

    \item \gls{fkf} algorithm, \eqref{fixed.variance} becomes
\begin{align}
\label{Regularized.NLMS}
    \bw_{t}&=\bw_{t-1} + \bx_t\frac{e_t}{v_\eta/\ov{v}+\|\bx_t\|^2 },
\end{align}
which is a regularized \gls{nlms} algorithm, with the regularization factor, $v_\eta/\ov{v}$, directly proportional to the variance of the noise $v_\eta$ and inversely proportional to the assumed variance, $\ov{v}$, of elements in $\btheta_t$.
    
    \item \gls{skf} algorithm, 
    produces
\begin{align}
\label{w.t.BKF}
    \bw_{t}&= \bw_{t-1} + \bx_t\frac{e_t}{v_\eta/\ov{v}_t+\|\bx_t\|^2 },
\end{align}
    where $\ov{v}_t$ is defined by \eqref{variance.update} for the \gls{skf} filter, and we recognize \eqref{w.t.BKF} to be the broadband Kalman filter \cite{Enzner10}, \cite[Sec. VII]{Paleologu13}, which generalizes the regularized \gls{nlms} in \eqref{Regularized.NLMS} by using the time-varying regularization factor $v_\eta/\ov{v}_t$.

    \item \gls{vkf} algorithm becomes
\begin{align}
\label{w.t.vKF.Gaussian.case}
    \bw_{t}&= \bw_{t-1} + \ov\bv_t\odot\bx_t\frac{e_t}{v_\eta+\ov\bv_t\T\bx_t^2 },
\end{align}
where $\bx_t^2=\bx_t\odot\bx_t$. It is easily seen to be a generalization of \eqref{w.t.BKF}. Note that $\ov\bv_t$ is calculated as shown in the first column of \eqref{variance.update} which makes \eqref{w.t.vKF.Gaussian.case} different from similar formulations that can be found in the literature, \eg \cite[Sec.~V]{Paleologu13}, where the vector $\ov\bv_k$ is calculated from previous estimates $\bw_{t-1}, \bw_{t-2}, \ld$.

\end{itemize}





\section{Generalized Gaussian noise and robust adaptive filters}\label{Sec:Generalized.Gaussian}

By robust filters, we understand those that can cope with non-Gaussian noise $\eta_t$, which essentially means that large-amplitude errors $\eta_t$ occur with a much higher probability that it would be the case of Gaussian noise (when keeping the variance of the noise constant).  

The idea is thus to use a generic non-Gaussian noise model and apply it to the general formulation of filtering, we presented in Sec.~\ref{Sec:Approximate.Kalman} and Sec.~\ref{Sec:Simplification.covariance}.

To this end, we assume that the noise follows the generalized Gaussian distribution with the shape parameter $\beta$, \ie the \gls{pdf} is defined as
\begin{align}
    \pdf_{\eta_t}(e) &=\frac{\beta}{2c\Gamma(1/\beta)}\exp\big[-(|e|/c)^\beta\big]\\
    &\propto \exp\Big[-\Big(\frac{|e|}{\sigma_\eta \kappa(\beta)}\Big)^\beta\Big],
\end{align}
where
\begin{align}
    \kappa(\beta)&=\sqrt{\frac{\Gamma(1/\beta)}{\Gamma(3/\beta)}}
\end{align}
and
\begin{align}
    \sigma_\eta = \sqrt{\Ex[\eta_t^2]}=\frac{c}{\kappa(\beta)}
\end{align}
is the standard deviation of $\eta_t$; thus, the variance is given by $v_{\eta}=\frac{c^2}{\kappa^2(\beta)}$. 

\begin{figure}[tb]
    \centering
    \includegraphics[width=1.0\linewidth]{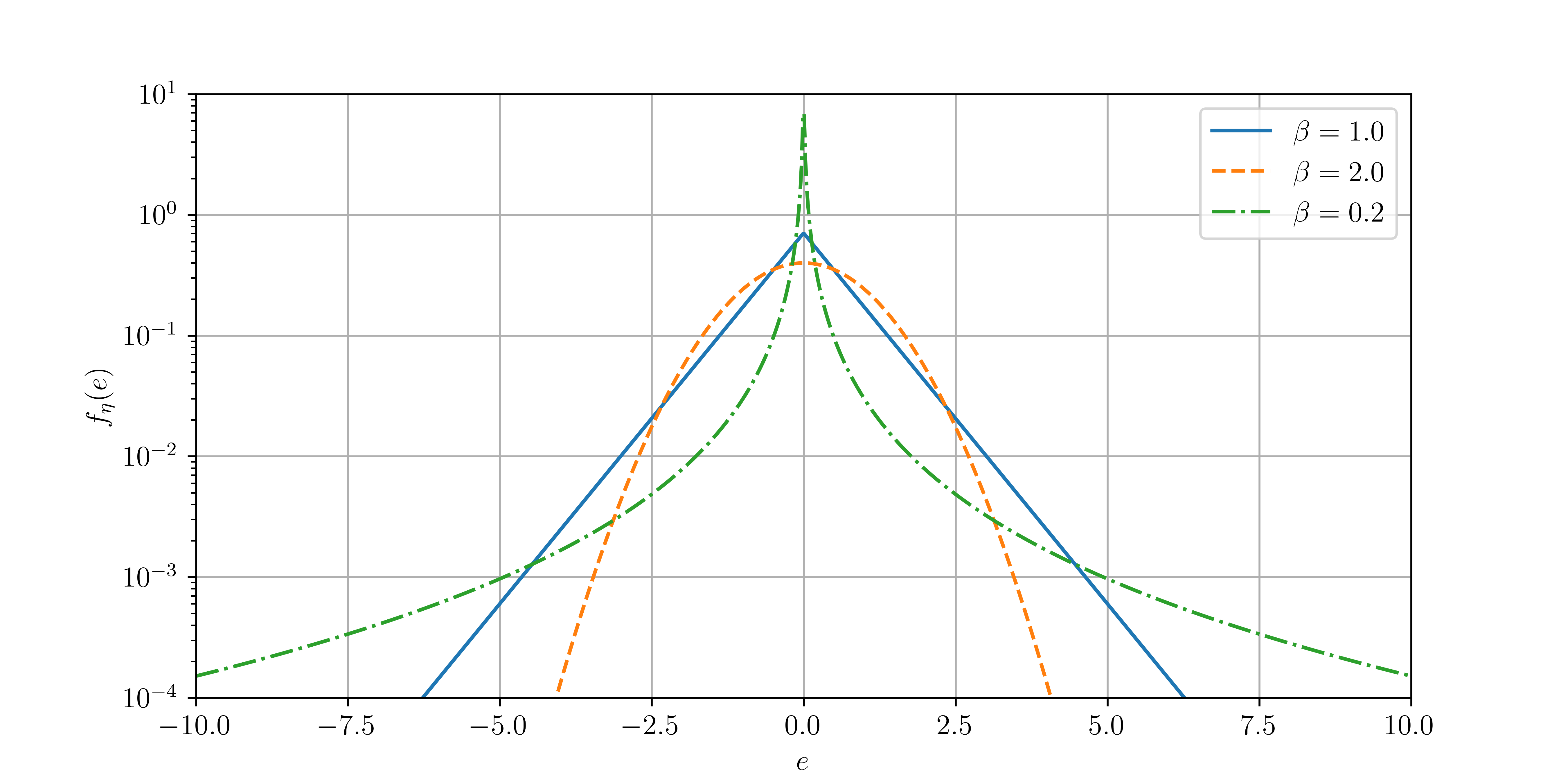}
    \caption{\gls{pdf}s of a generalized Gaussian variable with zero-mean and unit variance. By changing the shape parameters we obtain the Gaussian ($\beta=2.0$), the Laplace ($\beta=1.0$), and a heavy-tailed distributions ($\beta=0.2$).}
    \label{fig:log_PDF}
\end{figure}

With $\beta<1$, we obtain a ``heavy-tailed" distribution, where the ``outliers" have a large probability of occurrence; for $\beta=1$, we obtain the Laplace distribution; while when $\beta=2$, we return to the Gaussian distribution, with variance $v_{\eta}=c^2/2$. Examples of the \gls{pdf}s are shown in Fig.~\ref{fig:log_PDF}.

The likelihood \eqref{ell.definition} is then given by
\begin{align}
\label{ell.Gen.Gaussian}
    \ell(e)&=-\frac{|e|^\beta}{c^\beta}+ \tnr{Const.}
\end{align}
and, as we show in Appendix~\ref{App:Minorization}, the minorization function \eqref{ell(z).approx.quadratic} is defined by the term $h_t=h(e_t)$, where
\begin{align}
\label{h_t.Gen.Gaussian}
    h(e_t) &= \frac{\beta}{c^\beta} |e_t|^{\beta-2}=\frac{\beta}{[\sigma_{\eta} \kappa(\beta)]^\beta} |e_t|^{\beta-2};
\end{align}
the dependence on the prediction error $e_t$ can be appreciated, although it disappears in the Gaussian case, \ie $h_t=\frac{1}{v_\eta}$ for $\beta=2$.

We may now use \eqref{h_t.Gen.Gaussian} in the \gls{kf}, \gls{skf}, \gls{fkf}, or \gls{sg} algorithms, shown in Sec.~\ref{Sec:Simplification.covariance}, which will produce a new family of robust adaptive filters parameterized with $\beta\in[1, 2]$. For $\beta=2$ (Gaussian case), we recover the conventional solutions we have already shown at the end of Sec.~\ref{Sec:Simplification.covariance}. On the other hand, for $\beta<2$, we obtain robust adaptive filters that can be tailored to the distribution of measurement noise. 

\subsection{Robust adaptive filters} \label{Sec:Algorithms.General_case}

We can now use the previous formulation and specialize them to the expression in the generalized Gaussian case.

The \gls{sg} algorithm, \eqref{SG.algoritm}, becomes
\begin{align}
\label{SG.Gen.Gaussian}
    \bw_{t}
    &=\bw_{t-1} + \mu \bx_t  |e_t|^{\beta-1}\tnr{sign}(e_t) ,\\
    \mu &= \frac{\beta \ov{v}}{\big[\sigma_{\eta} \kappa(\beta)\big]^\beta},
\end{align}
with the step size depending not only on the variance of the noise $\eta_t$ and of the estimates $\btheta_t$ (\ie on $v_\eta=\sigma_\eta^2$ and $\ov{v}$), but also on the shape parameter $\beta$.

The \gls{fkf} algorithm, shown in \eqref{fixed.variance}, becomes
\begin{align}
\label{fixed.variance.Gen.Gaussian}
    \bw_{t}
    &=\bw_{t-1} + \bx_t\frac{|e_t|^{\beta-1}\tnr{sign}(e_t)}
    {\tau/\ov{v}+\|\bx_t\|^2/|e_t|^{2-\beta}}  \\
\label{fixed.variance.Gen.Gaussian.stable}
    &=\bw_{t-1} + \bx_t\frac{e_t}
    {|e_t|^{2-\beta}\tau/\ov{v}+\|\bx_t\|^2} , \\
\label{fixed.variance.Gen.Gaussian.tau}
    \tau
    &=\frac{\big(\sigma_{\eta} \kappa(\beta)\big)^\beta}{\beta} ,
\end{align}
where \eqref{fixed.variance.Gen.Gaussian.stable} is the numerically convenient representation, which has also an appealing interpretation: for $\beta\in(1,2)$, the regularization of the \gls{nlms} increases with the absolute value of the prediction error $|e_t|^{2-\beta}$.

The \gls{skf} algorithm becomes
\begin{align}
\label{v_bar.t.sKF.Gen.Gaussian}
    \ov{v}_{t}&= v_{t-1}+\varepsilon , \\
    \bw_{t}
    &=\bw_{t-1} + \bx_t\frac{|e_t|^{\beta-1}\tnr{sign}(e_t)}
    {\tau/\ov{v}_t+\|\bx_t\|^2/|e_t|^{2-\beta}}\\
    &=\bw_{t-1} + \bx_t\frac{e_t}
    {|e_t|^{2-\beta}\tau/\ov{v}_t+\|\bx_t\|^2} , \\
\label{v.t.sKF.Gen.Gaussian}
    v_{t}
    &=
    \ov{v}_{t} \left(1 - \frac{1}{M}
    \frac{  \|\bx_t\|^2}
    {|e_t|^{2-\beta}\tau/\ov{v}_t+\|\bx_t\|^2}\right). 
\end{align}

The \gls{vkf} and \gls{kf} algorithms require the gain multiplier $\alpha_t$ which is calculated as follows:
\begin{align}
\label{a.t.KF}
    \alpha_t 
    &=  \frac{|e_t|^{\beta-2}}{\tau + |e_t|^{\beta-2} s_t}\\
    &= \frac{1}{\tau |e_t|^{2-\beta} +s_t}.
\end{align}

\subsection{Sign-error algorithms} \label{Sec:Algorithms.signed.error}
Akin to the particular Gaussian noise case discussed in Sec.~\ref{Sec:Sanity.check.Gaussian.case}, we can now consider the particular case of $\beta=1$, which implies that the noise $\eta_t$ is Laplacian. By doing so
\begin{itemize}
    \item 
    \gls{sg} algorithm, \eqref{SG.Gen.Gaussian}, becomes
\begin{align}
\label{signed.LMS}
    \bw_{t}
    &=\bw_{t-1} + \mu \bx_t  \tnr{sign}(e_t) , \\
    \mu &= \frac{\ov{v}}{\sqrt{2}\sigma_{\eta}},
\end{align}
which we recognize as the sign-error LMS algorithm \cite[Ch.~12.1]{Sayed08_Book}.
    \item 
    \gls{fkf} algorithm, shown in \eqref{fixed.variance.Gen.Gaussian}-\eqref{fixed.variance.Gen.Gaussian.tau}, becomes
\begin{align}
\label{fixed.variance.Laplace}
    \bw_{t}
    &=\bw_{t-1} + \bx_t\frac{\tnr{sign}(e_t)}
    {\tau/\ov{v}+\|\bx_t\|^2/|e_t|}\\
    &=\bw_{t-1} + \bx_t\frac{e_t}
    {|e_t|\tau/\ov{v}+\|\bx_t\|^2},\\
\label{tau.beta=1}
    \tau
    &=\sigma_{\eta} \sqrt{2},
\end{align}
which is a regularized version of the normalized sign-error LMS algorithm, and is similar to the form shown in \cite[Eq.~(24)]{Deng05}.
    \item 
    \gls{skf} algorithm \eqref{v_bar.t.sKF.Gen.Gaussian}-\eqref{v.t.sKF.Gen.Gaussian} becomes
a variable step-size version of the regularized and normalized sign-error LMS algorithm:
\begin{align}
\label{v_bar.t.sKF.Gen.Gaussian.beta=1}
    \ov{v}_{t}&= v_{t-1}+\varepsilon , \\
    \bw_{t}
    &=\bw_{t-1} + \bx_t\frac{\tnr{sign}(e_t)}
    {\tau/\ov{v}_t+\|\bx_t\|^2/|e_t|} \\
    &=\bw_{t-1} + \bx_t\frac{e_t}
    {|e_t|\tau/\ov{v}_t+\|\bx_t\|^2} , \\
\label{v.t.sKF.Gen.Gaussian.beta=1}
    v_{t}
    &=
    \ov{v}_{t} \left(1 - \frac{1}{M}
    \frac{  \|\bx_t\|^2}
    {|e_t|\tau/\ov{v}_t+\|\bx_t\|^2}\right), 
\end{align}
with $\tau$ given by \eqref{tau.beta=1}.

    \item
    \gls{vkf} and \gls{kf} algorithms are obtained by using 
\begin{align}
\label{a.t.KF.beta=1}
    \alpha_t 
    &= \frac{1}{\tau |e_t| +s_t}
\end{align}
in \eqref{a_t.definition} or \eqref{w_t.update.iterations}. 
\end{itemize}

Note that, beside the sign-error LMS algorithm in \eqref{signed.LMS} (which gives name to this family of algorithms), all other algorithms are new. This is notable, as they are obtained almost effortlessly by applying the generic equations we derived in Sec.~\ref{Sec:Approximate.Kalman}.

\section{Numerical examples}\label{Sec:Numerical.examples}

\begin{figure}[tb]
    \centering
    \includegraphics[width=1.0\linewidth]{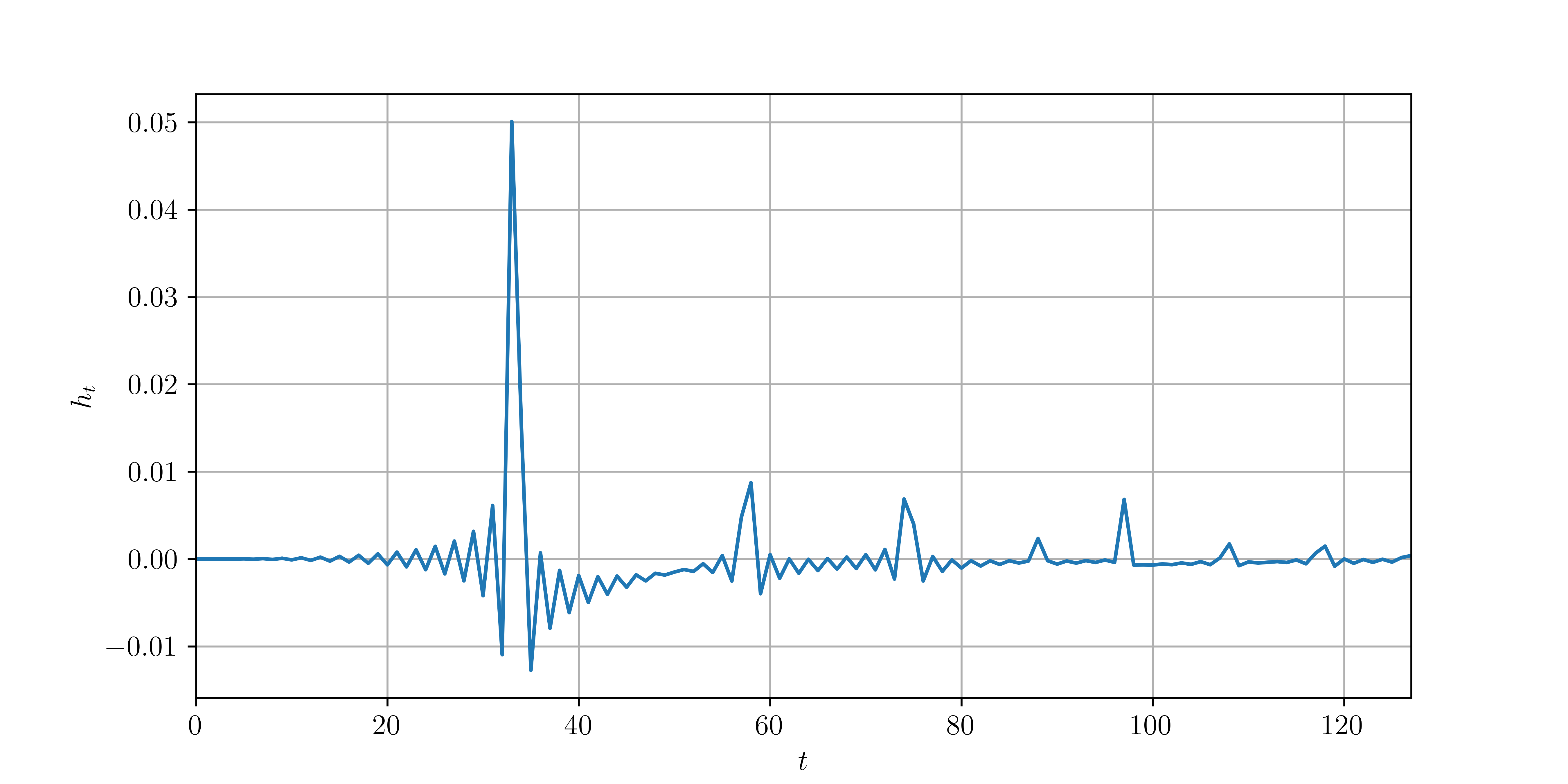}
    \caption{Impulse response, $h_t$, used in numerical examples.}
    \label{fig:h_t}
\end{figure}

We consider the scenario, where we observe the acoustic signals: output signal, $y_t$, and input signal, $x_t$, which are linearly related as
\begin{align}
\label{linear.convolution}
    y_t=\bx_t\T \bh + \eta_t,
\end{align}
where 
\begin{align}
    \bx_t &= \big[ x_{t}, x_{t-1},\dots, x_{t-M+1} \big]\T,
\end{align}
and
\begin{align}
    \bh &= [h_0, h_1,\ld h_{M-1}]\T
\end{align}
is the acoustic impulse response shown in Fig.~\ref{fig:h_t}, which for $M=128$ is calculated using the software in~\cite{Werner23} for a room of dimensions $(5, 10, 6)$~m, where the source is in position $(1, 2.5, 2)$~m and the receiver is in position $(1, 1.5, 1)$~m, with a sampling rate of $8$~kHz and a reverberation time of $200$~ms.

The input signal is generated from the autoregressive process $x_t=-a x_{t-1} + u_t$, where we use $a=0.9$ and $u_t$ is a zero-mean white Gaussian noise with variance $v_u$. Measurement noise $\eta_t$ is generated as a zero-mean, white generalized Gaussian variables with shape parameter $\beta^*$ and variance $v^*_\eta$.

We define the output \gls{snr} as
\begin{align}
\label{SNR.definition}
\tnr{SNR} & = 10\log_{10}\left\{ \frac{\Ex[|\bx_t\T \bh|^2]}{v^*_\eta}\right\}~[\tnr{dB}] ,
\end{align}
where $\Ex[\cd]$ is the expectation with respect to $x_t$. For any given $v_u$ and $\tnr{SNR}$, we thus find $v^*_\eta$ via \eqref{SNR.definition}.

Our goal is to identify the impulse response, $\bh$, and since the observation equation \eqref{output.equation.general} perfectly matches the model \eqref{linear.convolution}, we treat the result of adaptive filtering $\bw_t$, as estimates of $\bh$.

The estimation quality is assessed by the misalignment:
\begin{align}    
\label{Misalignement.def}
\mfm_t &= \frac{\|\bw_t - \bh\|^2_2}{\|\bh\|^2_2},
\end{align}
or, by its average:
\begin{align}    
\label{Avg.Misalignement.def}
\ov{\mfm}_t & = \Ex[\mfm_t],
\end{align}
where the expectation is taken over $x_t$ and $\eta_t$. In practice, this is done by averaging $\mfm_t$ over $N$ independent realizations of the noise $\eta_t$ and of the input signal $x_t$; here we use $N=100$.

Although the algorithms are derived assuming, in \eqref{state.equation.general}, that the state $\btheta_t$ varies in time, the impulse response, $\bh$, in our example is constant; so, with appropriately set algorithm parameters, the misalignment should converge to a useful/small value $\ov\mfm_\infty$. 

Thus, by choosing different parameters of the algorithms, we will trade the convergence rate against the misalignment at convergence $\ov{\mfm}_\infty$. For example, a larger step size $\mu$ in the \gls{sg} algorithm increases the rate at the cost of increased misalignment \cite[Ch.~5.4]{haykin2014}. 

To simplify the comparison, it is customary to assess convergence for algorithms with the same misalignment $\ov{\mfm}_\infty$, \cite{morgan1989}. Here, we do it by running the algorithm on a parameter grid and choosing those that reach the ``small'' target misalignment $\ov\mfm_\infty$ (\eg $\ov\mfm_\infty=0.01=-20$~dB).

To simplify the comparison of the algorithms, we assume that the variance of the noise $v^*_\eta$ is known (\eg estimated from $y_t$ in the absence of the input signal $x_t$) and, in the algorithms, we use $v_\eta=v^*_\eta$.\footnote{For details about the impact of an imperfect estimate of the measurement noise variance on the algorithms performance, see  discussions in \cite[Example~7]{kuhn2018stochastic} and \cite[Example 3]{becker2024npvss}.} However, we note that this information can be exploited in a meaningful way only in the \gls{skf}, \gls{vkf}, and \gls{kf} algorithms, where the Kalman gain $\bkappa_t$ (which depends on $v_\eta$) interacts, respectively, with $\ov{v}_t$, $\ov{\bv}_t$, or $\ov\matV_t$. On the other hand, the \gls{sg} and \gls{fkf} algorithms depend on the regularization factor defined by a ratio $\tau/\ov{v}$, where we must \emph{assume} a suitable variance $\ov{v}$. However, instead of assuming that we know $\tau$ and $\ov{v}$, we might assume that we know directly the regularization factor.

In the examples, we use non-Gaussian noise $\eta_t$, with $\beta^*=0.2$ while, for the adaptive algorithms, we consider two cases. The first is the ``conventional'' adaptive filtering, which assumes Gaussian noise, \ie uses $\beta=2.0$. The second case corresponds to a robust filter, which assumes Laplacian noise, \ie  is based on $\beta=1.0$, which, as mentioned before, is the smallest value of $\beta$ that guarantees that the function $\ell_y(z)$ is concave and thus the solution of \eqref{w_t=argmax.approx} is unique.

Therefore, in all cases we study, the adaptive filters are mismatched with respect to the distribution of the noise but, to approximate the heavy-tailed noise, we expect the Laplacian assumption ($\beta=1.0$) be more suitable than the Gaussian one ($\beta=2.0$).

As already mentioned in Sec.~\ref{Sec:Sanity.check.Gaussian.case}, in the conventional case, \ie for $\beta=2.0$, the \gls{sg} algorithm corresponds to the LMS algorithm, the \gls{fkf} algorithm is the same as the regularized \gls{nlms}, while the \gls{skf} corresponds to the broadband Kalman filter.

On the other hand, among robust filters, the \gls{sg} algorithm corresponds to the sign-error LMS, while the \gls{fkf}, \gls{skf}, and \gls{kf} algorithms are new members of the sign-error family of algorithms, see Sec.~\ref{Sec:Algorithms.signed.error}.

While we compare the performance of the algorithms, we emphasize that our goal is not to study their properties in detail. Rather, we want to show that, using the Bayesian framework, the algorithms can be almost effortlessly derived from the generic equations we have shown, and their performance decreases when the simplifications increase.

Finally, in numerical examples, the \gls{skf} and \gls{vkf} algorithms performed in a practically identical way; thus, to not overcrowd the presentation, we do not show the results for the latter. This is not to say that the \gls{vkf} algorithm is not useful and examples can be found in data processing, where the application of the \gls{vkf} algorithm provides advantages over the \gls{skf} algorithm; see \cite[Sec.~5.1.2]{Szczecinski21}.

\begin{example}[Non-iterative solutions]

The conventional adaptive filters results are shown in Fig.~\ref{fig:Heavytail.noise}a while the results obtained using the robust filters are presented in Fig.~\ref{fig:Heavytail.noise}b; we set $\ov\mfm_\infty=-20$~dB and use non-iterative versions, \ie $I=0$. For reproducibility purposes, the parameters used to obtain these results are shown in Table~\ref{Tab:parameters_algorithms.Heavytail}.

\begin{figure}
    \centering
\includegraphics[width=0.9\linewidth]{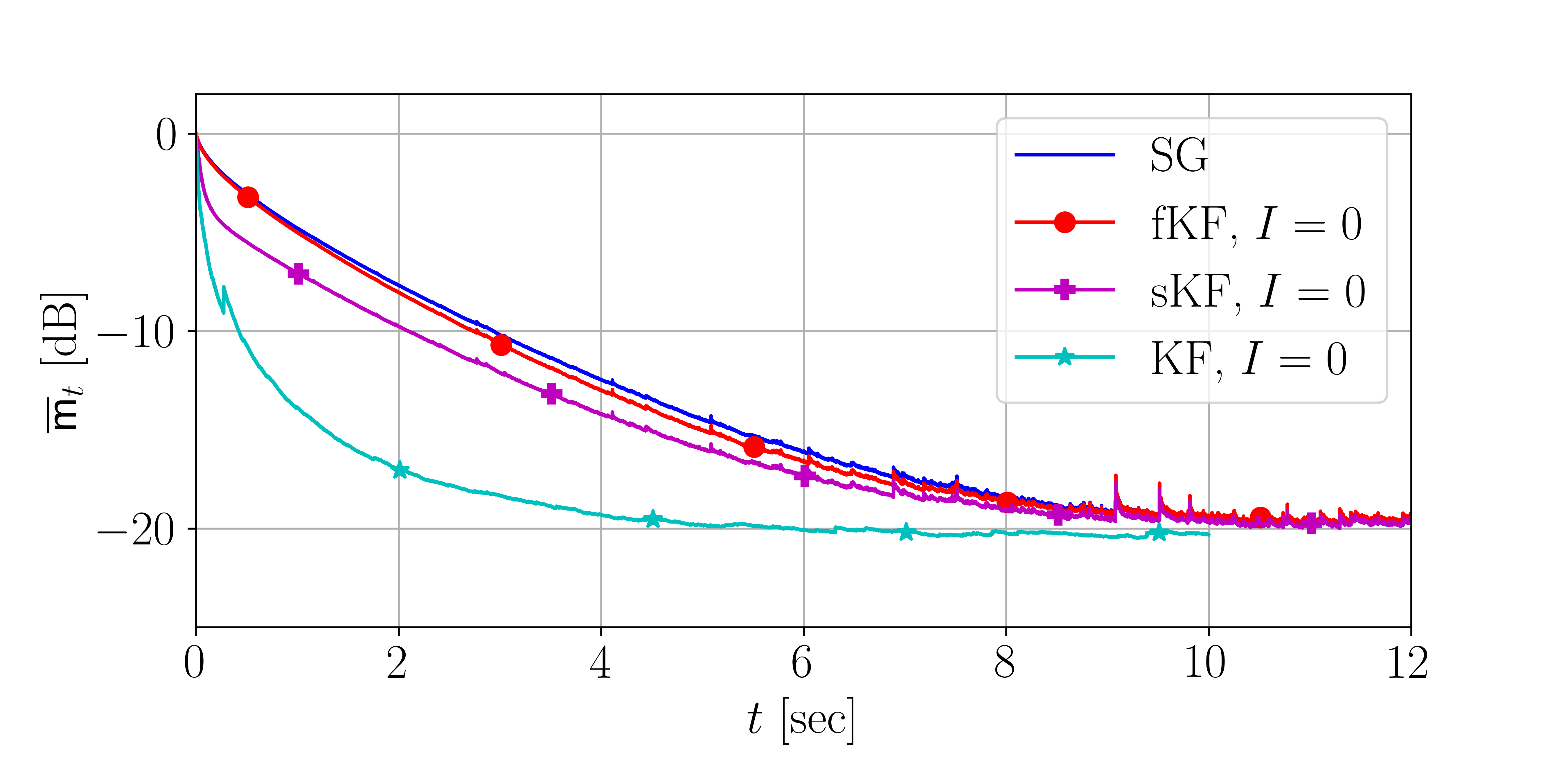}\\
a) Conventional adaptive filters, $\beta=2.0$, thus \gls{sg}=LMS, \gls{fkf} = regularized \gls{nlms}, \gls{skf} = Broadband Kalman filter \\   
\includegraphics[width=0.9\linewidth]{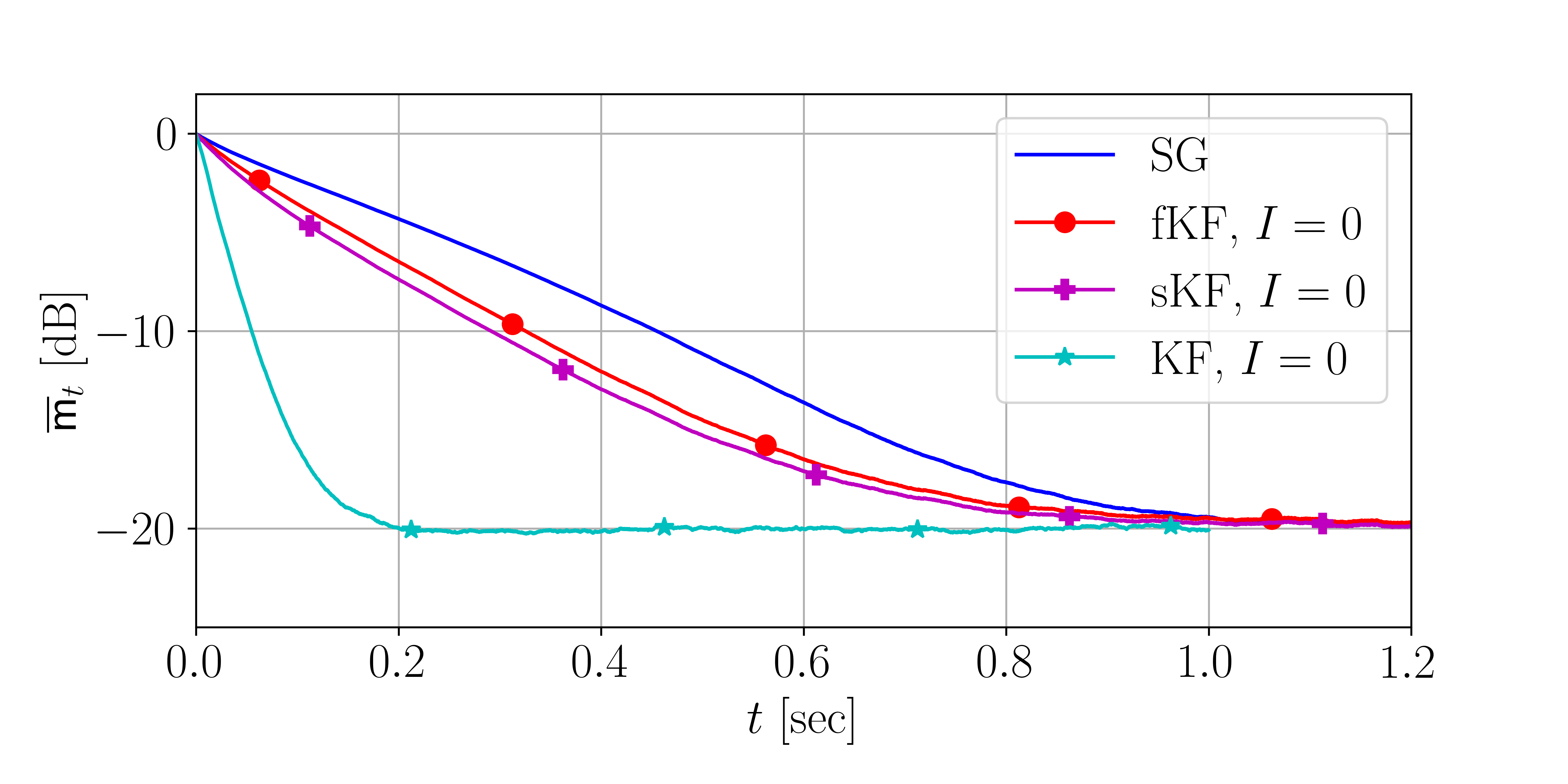}\\
b) Robust adaptive filters, $\beta =1.0$, thus \gls{sg} = sign-error LMS
    \caption{Convergence in heavy-tailed noise ($\beta^*=0.2$) for (a) conventional filters ($\beta=2.0$), and (b) robust filters ($\beta=1.0$); $\tnr{SNR} = 5$~dB and target misalignment $\ov\mfm_{\infty} = -20$~dB.}
    \label{fig:Heavytail.noise}
\end{figure}
    
\begin{table}[h]
    \centering
    \begin{tabular}{cc|l|l}
            & & \multicolumn{1}{c|}{$\beta=2.0$} & 
            \multicolumn{1}{c}{$\beta=1.0$} \\
        \hline
        SG & $\mu$  & $1.1\cd 10^{-4}$ & $2.7\cd 10^{-5}$\\
        \hline
        fKF & $\tau/\ov{v}$ & $8.2\cd 10^{3}$ & $1.1\cd 10^{4}$ \\
        \hline
        sKF & $\varepsilon$ & $3.2\cd 10^{-10}$ & $2.7\cd 10^{-8}$ \\
        \hline
        KF  & $\varepsilon$ & $ 3.6\cd 10^{-11}$ & $2.2\cd 10^{-8}$
    \end{tabular}
    \caption{Parameters of the adaptive algorithms used to obtain results shown in Fig.~\ref{fig:Heavytail.noise}.}
    \label{Tab:parameters_algorithms.Heavytail}
\end{table}

These results follow our intuition: by increasing the number of simplifying assumptions related to the dynamics of the filter, the convergence/performance of the algorithm deteriorates. This is less visible in conventional algorithms (see Fig.~\ref{fig:Heavytail.noise}a, where the \gls{fkf} and \gls{skf} algorithms perform similarly to the \gls{sg} algorithm) because the improvement in the model dynamics is tempered by the severe mismatch between the model of the noise (Gaussian) and the actual distribution (heavy-tailed). 

On the other hand, by using the robust filters, which better match the statistics of the measurement noise, the convergence improves notably: pay attention to the ten-fold difference in the time scales used in Fig.~\ref{fig:Heavytail.noise}a and in Fig.~\ref{fig:Heavytail.noise}b. Moreover, this also clarifies the gains due to improvement in model dynamics, where the \gls{sg} algorithm is clearly outperformed by the \gls{fkf} and \gls{skf} algorithms.

\end{example}

\begin{example}[Iterative calculation and variable target misalignment] 

In this example, we illustrate two effects: a) gains due to iterative recalculation of the Kalman gain multiplier $\alpha_t$ as defined in \eqref{w_t.update.iterations} (note that we only show the results for $I=1$ as the improvements observed when $I>1$ are negligible), and b) differences in convergence for different scenarios, which we enforce by changing the target misalignment $\ov\mfm_{\infty}\in\set{-15,-20,-25}$~dB. The parameters used to obtain these results are given in Table~\ref{Tab:variable.misalignment}.

\begin{table}[h]
    \centering
\scalebox{0.9}
{
    \begin{tabular}{cc|l|l|l|l|l|l}
            &  & \multicolumn{2}{c|}{$\ov\mfm_{\infty}=-15$~dB} & \multicolumn{2}{c|}{$\ov\mfm_{\infty}=-20$~dB} & \multicolumn{2}{c}{$\ov\mfm_{\infty}=-25$~dB}\\
            &  & $I=0$ & $I=1$ & $I=0$ & $I=1$ & $I=0$ & $I=1$\\
        \hline
        SG  & $\mu$ & $5.4\cd 10^{-5}$ & & 
        $2.7\cd 10^{-5}$ & & 
        $1.4\cd 10^{-5}$ \\
        \hline
        fKF & $\tau/\ov{v}$ & 
        $3.5\cd 10^{3}$ & $5.0\cd 10^{3}$ &
        $1.1\cd 10^{4}$ & $1.6\cd 10^{4}$ &
        $3.4\cd 10^{4}$ & $4.3\cd 10^{4}$\\
        \hline
        sKF & $\varepsilon$ & 
        $1.0\cd 10^{-7}$ & $7.1\cd 10^{-8}$ &
        $2.7\cd 10^{-8}$ & $2.2\cd 10^{-8}$ &
        $7.7\cd 10^{-9}$ & $6.6\cd 10^{-9}$\\
        \hline
        KF  & $\varepsilon$ & 
        $7.3\cd 10^{-8}$ & $6.0\cd 10^{-8}$ &
        $2.2\cd 10^{-8}$ & $2.2\cd 10^{-8}$ &
        $6.0\cd 10^{-9}$ & $6.0\cd 10^{-9}$
    \end{tabular}
}
    \caption{Parameters of the adaptive algorithms used to obtain the results shown in Fig.~\ref{Fig:variable.misalignment}.}
    \label{Tab:variable.misalignment}
\end{table}

\begin{figure}
    \centering
    \includegraphics[width=0.9\linewidth]{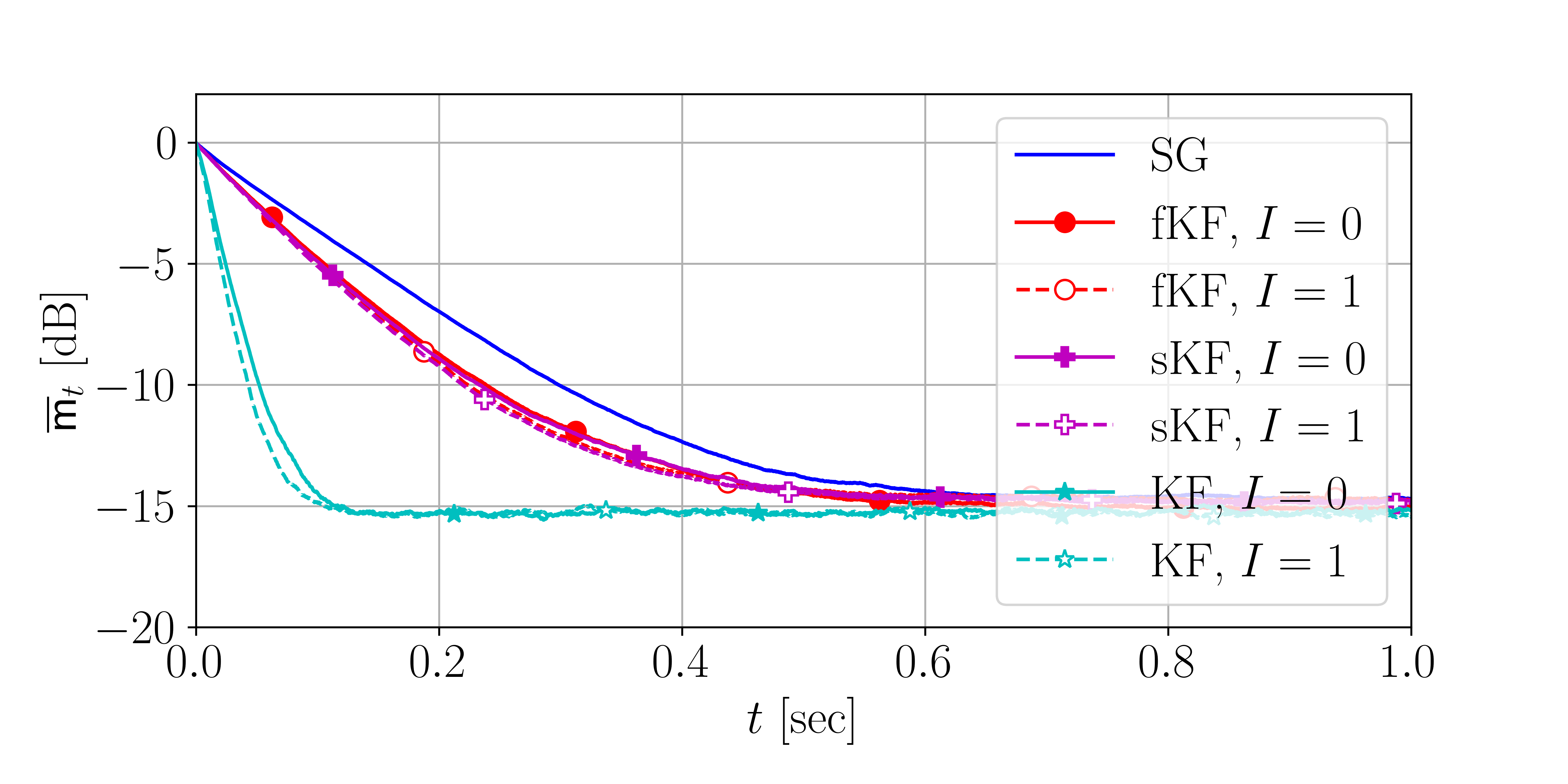}\\
    a) $\ov\mfm_{\infty}=-15$~dB\\
    \includegraphics[width=0.9\linewidth]{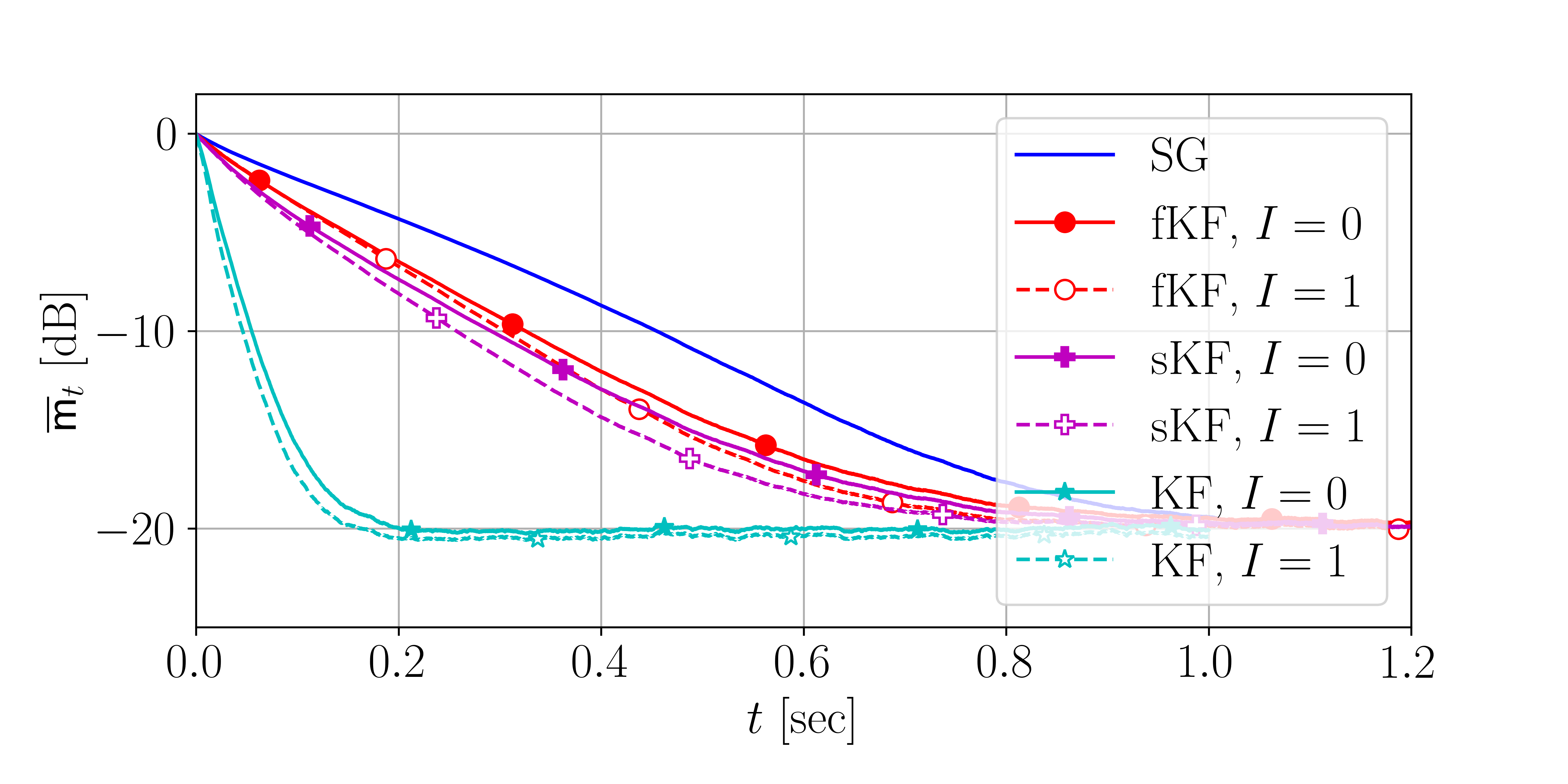}\\
    b) $\ov\mfm_{\infty}=-20$~dB\\
    \includegraphics[width=0.9\linewidth]{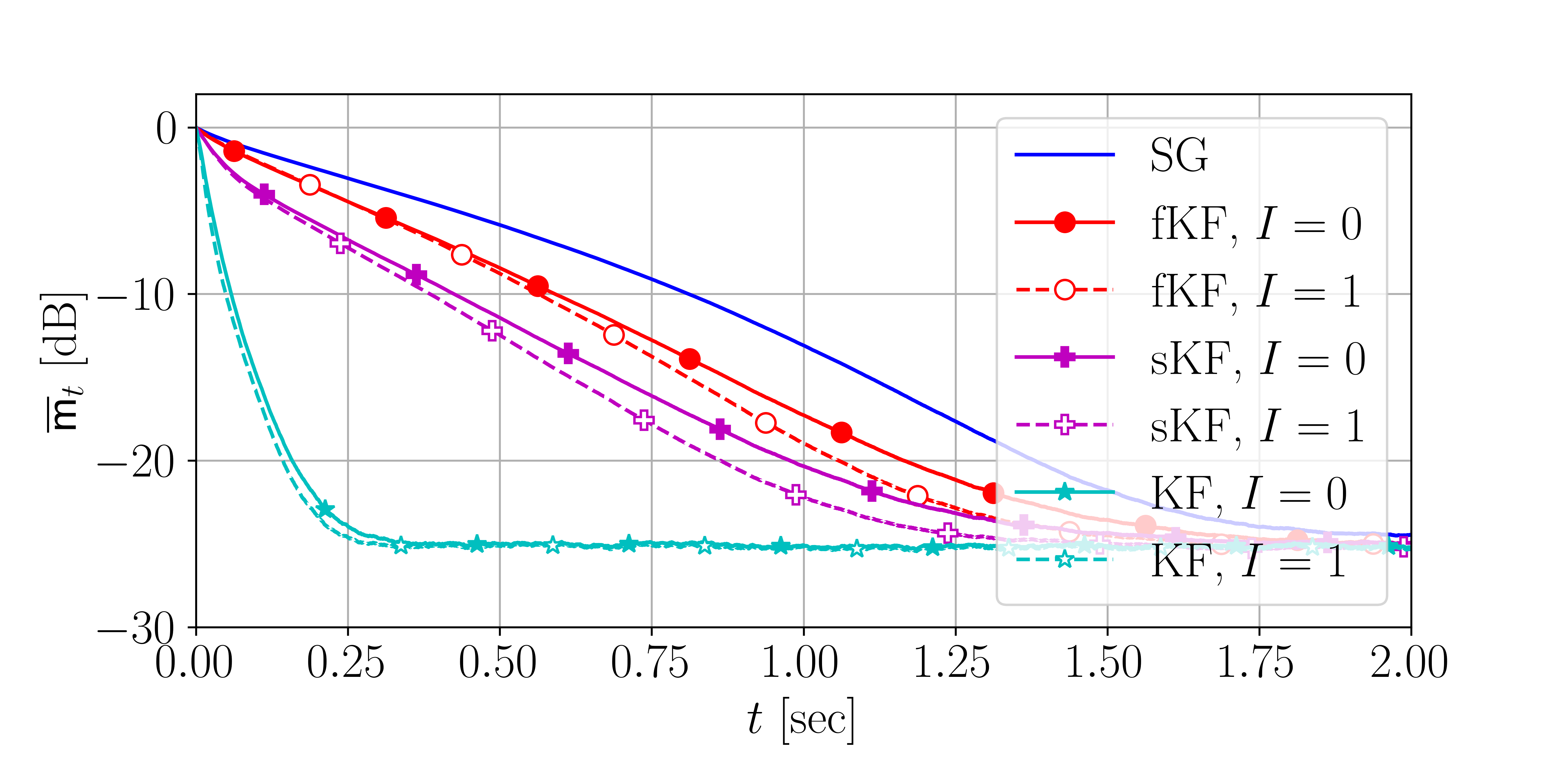}\\
    c) $\ov\mfm_{\infty}=-25$~dB\\
    \caption{Convergence of robust algorithms ($\beta=1.0$) in heavy-tailed noise ($\beta^*=0.2$) for target misalignment a) $\ov\mfm_{\infty}=-15$~dB, b) $\ov\mfm_{\infty}=-20$~dB, and c) $\ov\mfm_{\infty}=-25$~dB.}
    \label{Fig:variable.misalignment}
\end{figure}

To improve the performance (\ie decrease the misalignment), we obviously require more time to converge (note the change in the time-axis in Fig.~\ref{Fig:variable.misalignment}a--Fig.~\ref{Fig:variable.misalignment}c) and must adjust the parameters: decrease the adaptation step $\mu$ in the \gls{sg} algorithm, increase the regularization factor $\tau/\ov{v}$ in the \gls{fkf} algorithm, or decrease the variance $\varepsilon$ in the \gls{skf} and \gls{kf} algorithms.

These are rather well-known properties except the one referring to the regularization in the \gls{fkf} filter, where we note that the regularization is not a ``small'' value as it is sometimes stated to be \cite[Ch.~11]{Sayed08_Book}. In fact, being inversely proportional to the assumed variance $\ov{v}$, the regularization is also inversely proportional to the target misalignment $\ov\mfm_{\infty}$,\footnote{
This is because we may assume that $\bw_t$ are unbiased estimates of $\bh$, \ie $\Ex[\bw_\infty]=\bh$ (which, as we verified empirically, holds at the convergence) and thus the variance of the estimate of $\theta_{t,m}$ may be approximated as
\begin{align}
    \ov{v}_\infty  &\approx \frac{\ov\mfm_{\infty} \|\bh\|^2}{M}.
\end{align}
} 
which explains why the regularization grows when the latter decreases; in particular, in Table~\ref{Tab:variable.misalignment}, we see that the tenfold decrease of $\ov\mfm_{\infty}$ (from $-15$~dB to $-25$~dB) is obtained by a tenfold increase in the regularization value $\tau/\ov{v}$ (from $3.5\cd10^3$ to $3.4\cd10^4$).

The bottom line is that, by decreasing the target misalignment, we decide to operate in the small-variance (\ie small estimation error) regime, which accentuates the importance of accurate modeling, where the loss due to simplifications in the adaptive filters becomes more evident; indeed, we can observe a) the advantage of the \gls{skf} algorithm over the \gls{fkf} algorithm, and the advantage of the latter over the \gls{sg} filter, as well as, b) the added value of the iterative processing.

On the other hand, iterative recalculation of the Kalman gain multiplier $\alpha_t$ provides very modest improvement in the \gls{kf} filter. This should be attributed to the very fast convergence, where the gains of iterative refinement have no opportunity to materialize.
 
\end{example}

\section{Conclusions}\label{Sec:Conclusions}

In this work, using the concept of Bayesian filtering in the state-space model, we propose a unifying perspective for deriving conventional and robust adaptive filters. The following points are developed.
\begin{itemize}
    \item We explain how the well-known adaptive filters may be seen as particular cases of our Bayesian formulation if the Gaussian model of noise is adopted. Such a point of view explains, in a common framework, the origins of the well-known algorithms (such as the LMS, \gls{nlms}, regularized \gls{nlms}, and broadband Kalman filter), which are seen as simplified versions of the Kalman filter itself. 
        
    \item We apply the Bayesian formulation in more general (non-Gaussian) cases as well, which produces an entire family of adaptive filters depending on the parameters of the generalized Gaussian distribution.
    
    Here, beyond the general formulation, we also focus on the assumption of the Laplacian noise, for which we obtain a family of sign-error algorithms, which generalize the well-known sign-error LMS filter. Remarkably, these robust algorithms are derived effortlessly from the general Bayesian filtering formulations. This includes a robust Kalman filter, which is straightforwardly derived for non-Gaussian noise.
    
    \item We show that, due to the non-Gaussian nature of the noise, the filtering operation may be enhanced by iterative calculation of the Kalman gain. The validity of this proposition is verified by simulations and is shown to improve the performance when the adaptive filter targets low-error estimation.
\end{itemize}

We should note that although the relationship between the Kalman filter and other adaptive filters is not new \cite{Paleologu13}, it is not widely adopted. In fact, the literature more often associates the LMS filter with the approximate solution of the Wiener equations \cite[Ch.~10.1]{Sayed08_Book} and compares the LMS to the \gls{rls} \cite[Ch.~21.6]{Sayed08_Book}. While the latter can be related to the Kalman filter \cite[Ch.~31]{Sayed08_Book}, it requires a very specific transformation of the data and, in our view, is counterintuitive. Indeed, in the \gls{rls} formulation, the estimated parameters are deterministic, while in the \gls{kf} formulation they are random and must be tracked. In that perspective, the LMS algorithm, as well as other adaptive filters, which can be used for tracking, are more intuitively interpreted as approximate Kalman filters. 

Thus, beyond the derivation of new adaptive robust algorithms suited to operate in non-Gaussian noise, our formulation provides a unifying perspective on the adaptive filters. We believe that the possibility of deriving the well-known adaptive algorithms in a common framework is very useful from a pedagogical perspective and offers clear paths to derive new algorithms.

Moreover, by applying the simplifications, which yield the \gls{sg} and \gls{fkf} algorithms, we also obtain an insightful interpretation of the parameters (such as the adaptation step and the regularization factor). In particular, we dispel the myth that the regularization factor should be a ``small" value. Rather, we show that it is inversely proportional to the target estimation variance and, therefore, depending on the application, may be rather large. 
    
\begin{appendices}
\section{Minorization by quadratic function}\label{App:Minorization}

We start by removing the irrelevant constant term from $\ell(e)$ in \eqref{ell.Gen.Gaussian}, \ie we consider $\ell(e)=-\frac{|e|^\beta}{c^\beta}$. 

In addition, we exploit the property suggested in \cite[Sec.~ 3.4]{Hunter04a}, which states that $\ell(e)=-\frac{|e|^\beta}{c^\beta}$ can be minorized (\ie lower-bounded) by $q(e)=-\frac{1}{2}h_t e^2 + C_q$ if (i) $\ell(e_t)=q(e_t)$, (ii) $\ell'(e_t)=q'(e_t)$, and (iii) $\forall e , \ q''(e)\le\ell''(e)$. 

We focus on $e>0$ because both $\ell(e)$ and $q(e)$ are symmetric.

First, it is easy to show that the constant $C_q=(e_t/c)^\beta(\frac{1}{2}\beta-1)$ satisfies (i). Note that $C_q$ is irrelevant from an optimization perspective, but will be useful later in the proof.

The condition (ii) allows us to find $h_t=h(e_t)$ by making the first derivatives equal:
\begin{align}
    \ell'(e)|_{e=e_t}&=q'(e)|_{e=e_t} , \\
    -\beta\frac{|e_t|^{\beta-1}}{c^\beta}
    &=-h_t |e_t|,
\end{align}
which yields $h_t=\frac{\beta}{c^\beta} |e_t|^{\beta-2}$ shown in \eqref{h_t.Gen.Gaussian}.

We can now verify the condition (iii):
\begin{align}
    q''(e) &\le \ell''(e) ,\\
    -\frac{\beta}{c^\beta} e_t^{\beta-2} 
    &\le
    -\frac{\beta(\beta-1)e^{\beta-2}}{c^\beta} , \\
    e_t^{\beta-2} &\ge (\beta-1)e^{\beta-2},
\end{align}
which holds for $e>e_t$ and for all $\beta<2$. Moreover, if $\beta\le 1$, it also holds for any $e$.

If $\beta\in(1,2)$ and $0<e<e_t$, we proceed by definition, writing explicitly the minorization condition:
\begin{align}
    q(e)&\le \ell(e) ,\\
    -\frac{1}{2}\beta e_t^{\beta-2}e^2
    +e_t^\beta(\frac{1}{2}\beta-1)
    &\le 
    -e^\beta , \\
\label{u(x).p(x)}
    u(x)=x^\beta(\frac{1}{2}\beta-1)
    &\le 
    \frac{1}{2}\beta x^{\beta-2}-1=p(x),
\end{align}
where we set $x=e_t/e\ge1$. 

Both sides of inequality \eqref{u(x).p(x)} are equal for $x=1$, \ie $u(1)=p(1)$, and $u(x)$ is concave (for $\beta<2$) while $v(x)$ is convex, \ie $u''(x)<0$ and $p''(x)>0$. Thus, $u(x)$ is upper-bounded by $u(1)+u'(1)(x-1)$ while $p(x)$ is lower-bounded by $p(1)+p'(1)(x-1)$. Since $u'(1)<p'(1)$, the inequality is satisfied for the bounds when $x>1$, and thus, it must hold for the functions as well.

\begin{figure}[tb]
    \centering
    \includegraphics[width=1.0\linewidth]{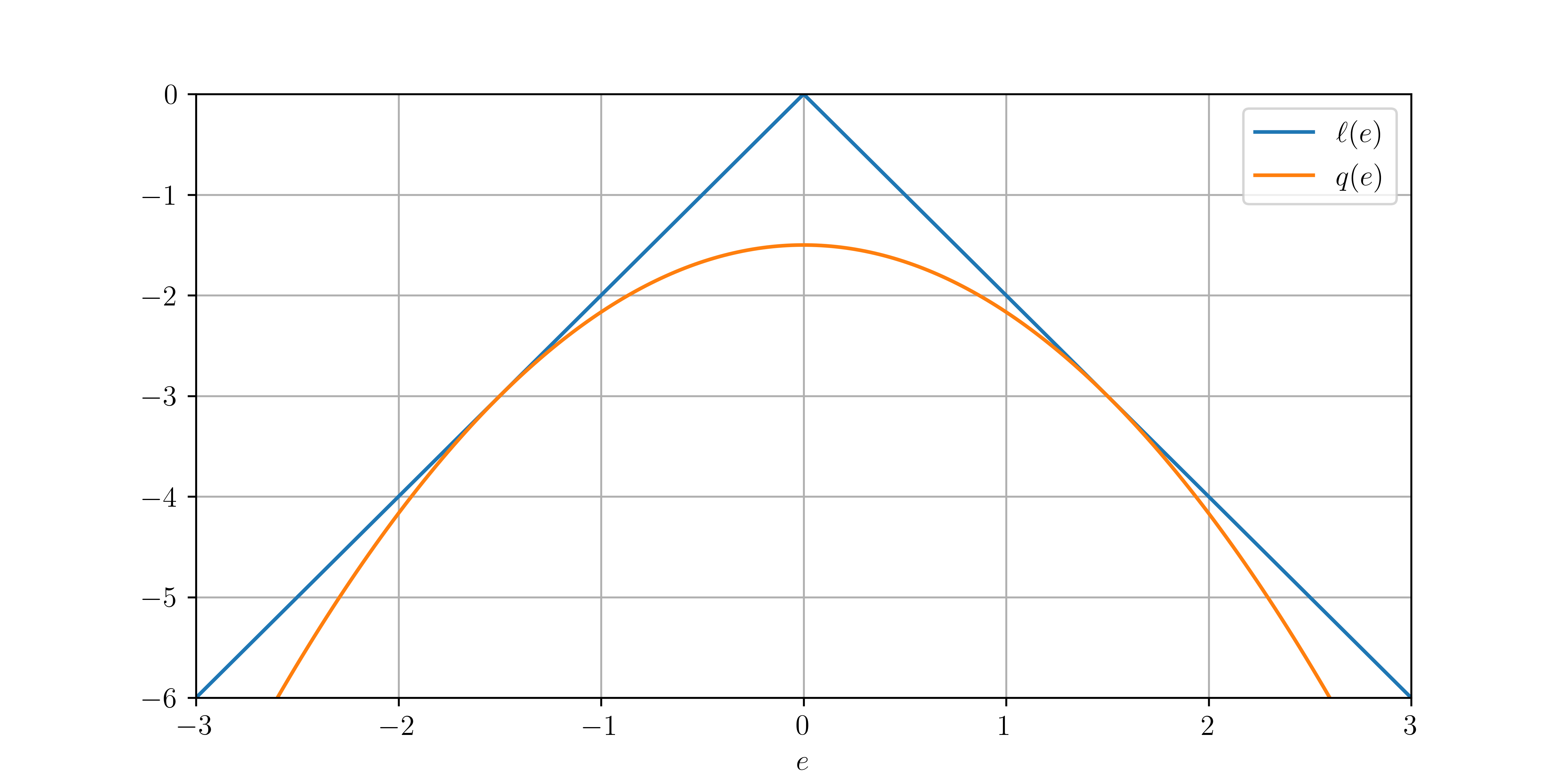}
    \caption{Minorization example for $\beta=1$ and $e_t=1.5$, which is the value where $\ell(e_t)=q(e_t)$.}
    \label{fig:minorization}
\end{figure}

An example of the minorization for $\beta=1$ is shown in Fig.~\ref{fig:minorization}.

\end{appendices}

\ifdefined\ARXIV

\bibliographystyle{\CFilesBib/IEEEtran}
\else
\bibliography{\CFilesBib/IEEEabrv,\CFilesBib/references_rank,\CFilesBib/references_all}
\bibliographystyle{\CFilesBib/IEEEtran}
\fi

\end{document}